\documentclass[oneside,a4paper,english]{article}
\usepackage[a4paper,left=1in,right=1in,top=1in,bottom=1in]{geometry}
\usepackage{graphicx}
\usepackage{amsmath,amsthm,amsfonts}
\usepackage{multirow}
\usepackage{url}
\usepackage{booktabs}
\usepackage{color}

\newenvironment{nalign}{
	\begin{equation}
	\begin{aligned}
}{
	\end{aligned}
	\end{equation}
	\ignorespacesafterend
}

\newenvironment{nalign*}{
	\begin{equation*}
	\begin{aligned}
}{
	\end{aligned}
	\end{equation*}
	\ignorespacesafterend
}

\newcommand{\parder}[2]{\frac{\partial #1}{\partial #2}}

\newcommand{\parderD}[2]{\frac{\mbox{D} #1}{\mbox{D} #2}}
\newcommand{\parderDtwo}[2]{\frac{\mbox{D}^2 #1}{\mbox{D} {#2}^2}}
\newcommand{\hzero}{H^1_0\left(\Omega\right)}

\newcommand{\honedual}{H^{-1}\left(\Omega\right)}
\newcommand{\sobz}[2]{W_0^{#1,#2}\left(\Omega\right)}
\newcommand{\sob}[2]{W^{#1,#2}\left(\Omega\right)}
\newcommand{\cont}{C\left(\bar{\Omega}\right)}
\newcommand{\norm}[2]{\left\lVert#1\right\rVert_{#2}}
\newcommand{\Sp}[1]{\textnormal{Sp}\left(#1\right)}
\newcommand{\Vp}[1]{\textnormal{Vp}\left(#1\right)}

\newcommand{\av}[1]{\left\langle#1\right\rangle}
\newcommand{\mueff}{\mu_{\scriptsize{\textnormal{eff}}}}

\newcommand{\coloremph}{black}

\newtheorem{theorem}{Theorem}
\newtheorem{lemma}{Lemma}
\newtheorem{proposition}{Proposition}
\newtheorem{remark}{Remark}
\newtheorem{definition}{Definition}

\numberwithin{equation}{section}

\DeclareMathOperator*{\essinf}{ess-inf}

\graphicspath{{figs/}}

\title{Analysis of a cavitation model including bubbles in thin film lubrication}

\author{Alfredo Jaramillo$^a$, Guy Bayada$^b$, Ionel Ciuperca$^c$, Mohammed Jai$^b$\\
{\small
(a) Instituto de Ci\^encias Matem\'aticas e de Computa\c{c}\~ao,
Universidade de S\~ao Paulo, 13560-970 S\~ao Carlos, Brazil
}\\
{\small
(b) Universit\'e de Lyon, CNRS, INSA de Lyon, Institut Camille Jordan UMR 5208, F-69621 Villeurbanne, France
}\\
{\small
(c) Universit\'e de Lyon, CNRS, Institut Camille Jordan UMR 5208, F-69622 Villeurbanne, France
}
}

\begin{document}
\vspace{3cm}

\maketitle


\begin{abstract}
In the lubrication area, which is concerned with thin film flow, cavitation has been considered as a fundamental element to correctly describe the characteristics of lubricated mechanisms. Here, the well-posedness of a cavitation model that can explain the interaction between viscous effects and micro-bubbles of gas is studied. This cavitation model consists of a coupled problem between the compressible Reynolds PDE (that describes the flow) and the Rayleigh-Plesset ODE (that describes micro-bubbles evolution). This coupled model  seems never to be studied before from its mathematical aspects. Local times existence results are proved and stability theorems are obtained based on the continuity of the spectrum for bounded linear operators. Numerical results are presented to illustrate these theoretical results. 

\medskip


\end{abstract}

\section{Introduction}\label{sec:introduction}
	Cavitation is observed in various engineering devices, ranging from hydraulic systems to turbo pumps for space applications. It is a challenging issue linked with various phenomenon: acoustic, thermodynamic and fluid dynamics. In the lubrication area, which is concerned with thin film flow, cavitation has been considered as a fundamental element to correctly describe the characteristics of lubricated mechanisms	 \cite{dowson1979,braun2010}. Cavitation has often been primarily associated with a diminution of the pressure $p$ in the liquid falling below the vapor pressure. Numerous models have been introduced to couple this unilateral condition with the Reynolds equation, which is usually used to model the pressure evolution  in thin film flow. Mathematical studies of these  models can be found in \cite{Cimatti1976,Cimatti1980,kinderlehrer1980,bayada1983,Bermudez1989,buscaglia2015a} in which existence and uniqueness results are given for both the stationary and transient cases. Another approach has been proposed in \cite{Bayada2006} by considering cavitation as a multifluid problem with a free boundary between two immiscible fluids. However, it is physically recognized that the cavitation phenomenon is linked with the existence and evolution of micro-bubbles in a liquid. This aspect has not been taken into account in these models.  It is however used in the well-known software Fluent for fluid mechanics \cite{schnerr2001,singhal2001,zwart2004} in which micro bubbles evolution is coupled  with the Navier Stokes system  for a 3-dimensional flow. In the lubrication area, this phenomenon has been ignored until the works  of Someya's group \cite{kawase1985,natsumeda1987} who proposed to couple the full Rayleigh-Plesset equation (which describes the evolution of a bubble) with the Reynolds equation (which describes the fluid). Numerous works follow in the lubrication literature using simplified forms of the Rayleigh-Plesset equation for various kind of applications \cite{kubota1992,gehannin2009,geike2009a,geike2009b,Jaramillo2019,Jaramillophd}. The paper of Snyder et al. \cite{snyder2016} can be considered as a review paper in this field.
	
	\subsubsection*{The Reynolds-Rayleigh-Plesset coupling}
	
\begin{figure}[h]\centering
		\includegraphics[width=.6\linewidth]{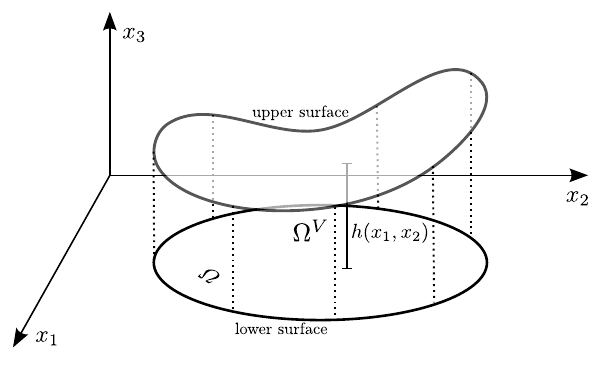}
		\caption{Three-dimensional scheme of the physical framework.}\label{fig:scheme3D}
	\end{figure}

		The fluid is contained in a domain $\Omega^V\subset\mathbb{R}^3$, limited by a domain $\Omega\subset\mathbb{R}^2$ in the $x_1$-$x_2$ plane, an upper surface given by the gap function $h\left(x_1,x_2\right)$ defined on $\Omega$ and by a vertical lateral boundary as shown in Fig. \ref{fig:scheme3D}. The surfaces are in relative movement along the $x_1$-$x_2$ plane at velocity $\mathbf{U}\in\mathbb{R}^2$. It is also assumed that the relative speed of the surfaces along the $x_3$-axis is null. In this work theoretical results on the well-posedness of the Reynold-Rayleigh-Plesset (RRP) cavitation model for the flow of a fluid multicomponent mixture are presented. Here, a brief description of that mathematical model is given, the physical hypotheses and a heuristic justification are given in the Appendix.
		
		The mixture is composed by two phases: an incompressible liquid phase (with known density $\rho_\ell$ and viscosity $\mu_\ell$) and a gas phase (with known density $\rho_g$ and viscosity $\mu_g$). It is assumed that the gas phase is composed by a distribution of bubbles immersed on the mixture, and that around a point $\mathbf{x}$ at time $t$ there can be bubbles of only one certain radius $R\left(\mathbf{x},t\right)$, i.e., the radii distribution is \emph{monodisperse}. In addition, the dynamics of the field $R\left(\mathbf{x},t\right)$ are governed by the Rayleigh-Plesset equation (e.g., \cite{brennen1995}):
\begin{equation}
\rho_\ell\left[\frac{3}{2}\left(\parderD{R}{t}\right)^2+R\parderDtwo{R}{t}\right]=P_0\left(\frac{R_0}{R}\right)^{3k}-\left(\bar{p}+p_\partial\right)-\frac{2\sigma}{R}-4\left(\frac{\mu_\ell+\kappa^s/R}{R}\right)\parderD{R}{t},\label{eq:rayleigh-plesset-full}
\end{equation}
where the terms at the left hand side are called \emph{inertial terms}, $\bar{p}$ is the averaged mixture's pressure, $P_0$ is the inner pressure of the bubble when its radius is equal to $R_0$, $k$ is the polytropic exponent {\color{\coloremph}(see the Appendix)}; $\sigma$ is the surface tension, $\kappa^s$ is the surface dilatational viscosity \cite{snyder2016}; $p_\partial$ is the pressure at the boundary; and
\begin{equation}
	\parderD{}{t} = \parder{}{t}+\left(\mathbf{{u}}_b\cdot \nabla\right),
\end{equation}
with $\mathbf{{u}}_b\in\mathbb{R}^3$ corresponding to the transport velocity of the bubbles. In the right-hand side of Eq. \eqref{eq:rayleigh-plesset-full} the first term models the pressure of the gas contained in the bubble.

In this work the transport velocity of the bubbles is assumed to be null, $\mathbf{u}_b=0$. This hypothesis covers cases where the bubbles are attached to one of the surfaces (that would be in relative motion), and cases where the surface's relative motion on the $x_1$-$x_2$ plane is null (for instance Pure Squeeze problems, e.g., \cite{gehannin2009,Gehannin2016,geike2009a,geike2009b}). Then, equation \eqref{eq:rayleigh-plesset-full} relates two unknown fields: $R$ and $\bar{p}$. A second equation is obtained by introducing the local gas fraction ($\frac{\textnormal{volume of gas}}{\textnormal{total volume}}$) in terms of $R$:
\begin{equation}
	\alpha=\alpha\left(R\left(\mathbf{x},t\right)\right)\label{eq:alphaintro},
\end{equation}
and relating the averaged mixture density $\bar{\rho}$ to $\alpha$ by means of (e.g., \cite{Drew1999}):
\begin{equation}
\bar{\rho}\left(R\left(\mathbf{x},t\right)\right)=\rho_g\,\alpha\left(R\left(\mathbf{x},t\right)\right)  + \rho_\ell\left(1-\alpha\left(R\left(\mathbf{x},t\right)\right)\right).
\end{equation}
It is also assumed some model for the mixture effective viscosity field, denoted $\mu_{\scriptsize{\textnormal{eff}}}$, in terms of the gas fraction $\alpha$, so one may write $\mu_{\scriptsize{\textnormal{eff}}}=\mu_{\scriptsize{\textnormal{eff}}}(\alpha(R))$. The RRP model assumes the averaged mixture pressure $\bar{p}$ accomplishes the compressible Reynolds equation:

\begin{equation}
\nabla_x\cdot \left(\frac{\bar{\rho} h^3}{12\mueff}\,\nabla \bar{p}_\ell\right)=\nabla_x\cdot \left(\frac{\mathbf{U}}{2}\,\bar{\rho} h\,\right)+\parder{\bar{\rho}h}{t}\qquad\textnormal{in }\Omega.\label{eq:Reynolds}
\end{equation}

The RRP cavitation model consists in the coupling of Eqs. \eqref{eq:rayleigh-plesset-full} and \eqref{eq:Reynolds} along with suitable boundary conditions.

	It is noteworthy that there exist many works in Mechanics' literature concerning the numerical resolution and modeling aspects of the coupling of the Rayleigh-Plesset equation with fluid flow equations (e.g., \cite{liuzzi2012,schmidt2014,liu2014,walters2015,zhao2016,dhande2016}). The well-known software FLUENT for Fluid Mechanics uses also this type of modeling \cite{schnerr2001,singhal2001,zwart2004}. On the other hand, in the mathematical field few works are concerned with this problem. The Rayleigh-Plesset equation alone without coupling (in which the pressure is a known data) has been subject of interest as differential equations with singularities \cite{hakl2012,ohnawa2016}. However, to the knowledge of the authors, no mathematical analysis of the full coupling of the Rayleigh-Plesset equation with a flow equation (Euler, Stokes or Reynolds) so far appeared.
	
	{\color{\coloremph} In this work a mathematical analysis is carried on by first writing an abstract form of the coupling RRP by means of auxiliary functions that depend on the unknown radii field $R$. Then, some general properties of these auxiliary functions are identified from the physics and held as hypotheses (e.g., positiveness, monotonicity, existence of critical points). Informally, a first step of the study consists in writing the coupled model as an ordinary differential equation on a Banach Space and making use of a suitable version of the Cauchy-Lipschitz Theorem. For this, it is shown that the unknown $p$ can be \emph{eliminated} by writing it in terms of $R$ and its derivatives. A second step regards the well-posedness of the stationary problem: the existence of a trivial stationary solution is established and then non-trivial solutions are found by continuity arguments; finally, continuity arguments are also used to extend the stability of the trivial solution to the stability of non-trivial cases. These two analysis steps are independently performed for two scenarios: 1) including or 2) disregarding the inertial terms in the Rayleigh-Plessset equation.} 
	
	The structure of this document is as it follows: after the introduction section, the mathematical framework is described in Section \ref{sec:math-framework} where notations and some previous required results are given. Section \ref{sec:mathematical-analysis-with-inertial-terms} is devoted to the study of the full system \eqref{eq:Reynolds} to \eqref{eq:rayleigh-plesset-full} including inertial terms; existence of a stationary solution is gained by way of the Implicit Function Theorem around some particular data for which a stationary solution is easy to compute; a stability result is obtained with a small data assumption by studying the spectrum of a differential operator, and the continuity of that spectrum around the particular data; at last, an instability result is gained in the one dimensional case by means of the Routh-Hurwitz Theorem. In Section \ref{sec:mathematical-analysis-without-inertial-terms}, a simplified Rayleigh-Plesset equation neglecting the inertial terms is considered; unlike the previous section, existence of the (local) solution of the system is not obvious and requires to use the Freedholm Alternative Theorem; stability results of the stationary solution for small data are obtained using also the spectrum's continuity of a differential operator. In Section \ref{sec:numerical-analysis} some numerical examples are shown {\color{\coloremph}where time convergence towards stationary solutions is observed}. Some topics on possible future work are mentioned in Section \ref{sec:future-work}. Finally, a heuristic justification of both Eq. \eqref{eq:Reynolds} and the RRP coupling is given in the Appendix.


\section{Mathematical framework}
\label{sec:math-framework}

In this section we introduce some notations and previous results to be used along this document. 

Let $\Omega\subset \mathbb{R}^N$, $N=1,2$ be a regular domain, and introduce the change of variables
{\color{\coloremph}\begin{equation}
p=\bar{p}_\ell(x,t)/\rho_\ell~,
\end{equation}}
we consider the abstract problem of finding $p(x,t)$, $R(x,t)>0$, with $x\in \Omega$ and $t\geq 0$, such that
\begin{equation}
\frac{3}{2}\frac{1}{R}\left(\parder{R}{t}\right)^2+\frac{\partial^2 R}{\partial t^2}=\frac{f_1\left(R\right)-{p}}{R}-\parder{R}{t}f_2\left(R\right)\label{eq:abstract-rp}
\end{equation}
and
\begin{align}
\begin{split}
\nabla_x\cdot \left(f_3\left(R\right) h^3\,\nabla {p}\right)&=\nabla_x\cdot \left(f_4\left(R\right)\mathbf{U}\,h\,\right)+h\,f_5\left(R\right)\parder{R}{t},\\
p&=0 \qquad\mbox{on } \partial \Omega.
\end{split}
\label{eq:abstract-reynolds}
\end{align}
where $\mathbf{U}\in\mathbb{R}^N$. Along the initial conditions for every $x\in  \bar{\Omega}$:
\begin{align}
\begin{split}
R(x,0)&=r_1(x), \\
\parder{R}{t}(x,0)&=r_2(x)
\end{split}\label{eq:abstract-cond}
\end{align}
and $r_1$, $r_2$ are regular known functions. The terms in the left hand side of Eq. \eqref{eq:abstract-rp} are named \emph{inertial terms}. In the next sections we study the wellposedness of problem \eqref{eq:abstract-rp}-\eqref{eq:abstract-reynolds}-\eqref{eq:abstract-cond} when including or disregarding the inertial terms.

For $\alpha,\beta\in \mathbb{R}$ with $\alpha<\beta$ we define:
$$B_{\alpha,\beta	}=\left\{w\in L^\infty\left(\Omega\right):\alpha \leq w \leq \beta\mbox{ a.e. on } \Omega \right\}.$$
We make also the following hypotheses:
\begin{itemize}
	\item[H1:] $f_1\in C^2\left(\mathbb{R}^+_*;\mathbb{R}\right)$, $\exists \bar{R},\delta_1\in\mathbb{R}^+_*$ such that $f_1\left(\bar{R}\right)=0$ and $f'_1\left(R\right)<0$ $\forall R \in [\bar{R}-\delta_1,\bar{R}+\delta_1]$. We denote $m_1=\underset{R\in [\bar{R}-\delta_1,\bar{R}+\delta_1]}{\min}\left|
	f'_1\left(R\right)\right|$ and $M_1=\underset{R\in [\bar{R}-\delta_1,\bar{R}+\delta_1]}{\max}\left|f'_1\left(R\right)\right|$;
	\item[H2:] $f_2\in C^2\left(\mathbb{R}_*^+;\mathbb{R}^+_*\right)$;
	\item[H3:] $f_3\in C^2\left(\mathbb{R}_*^+;\mathbb{R}\right)$ and $\exists m_3,M_3>0$ such that $m_3\leq f_3\left(r\right)\leq M_3$ $\forall r\in \mathbb{R}^+$;
	\item[H4:] $f_4\in C^2\left(\mathbb{R}_*^+;\mathbb{R}^+\right)$, $f_4'\left(r\right)<0$ $\forall r>0$;
	\item[H5:] $f_5\in C^2\left(\mathbb{R}_*^+;\mathbb{R}^-_*\right)$;
	\item[H6:] $h \in B_{m_0,M_0}$ for $0<m_0<M_0$ constants. We denote $h_0=\underset{\Omega}{\essinf}~h$.
\end{itemize}
\begin{remark}
The physical model given by Eqs. \eqref{eq:Reynolds} to \eqref{eq:rayleigh-plesset-full} is a particular case of problem \eqref{eq:abstract-rp} to \eqref{eq:abstract-cond} for which
\begin{equation}\color{\coloremph}
f_1\left(R\right)=\frac{1}{\rho_\ell}\left(P_0\left(\frac{R_0}{R}\right)^{3k}-p_\partial-\frac{2\sigma}{R}\right)\label{eq:def-f_1},
\end{equation}
\begin{align*}\color{\coloremph}
f_2\left(R\right)&=\frac{4}{\rho_\ell}\left(\frac{\mu_\ell+\kappa^s/R}{R^2}\right),&\color{\coloremph}f_3\left(R\right)&=\frac{1}{12}\frac{(1-\alpha\left(R\right))+\alpha\left(R\right)\rho_g/\rho_\ell}{(1-\alpha\left(R\right))\mu_\ell+\alpha\left(R\right)\mu_g},\\ \color{\coloremph}f_4\left(R\right)&=\frac{1}{2}\left[1+\alpha\left(R\right)\left(\rho_g/\rho_\ell-1\right)\right],&f_5\left(R\right)&=f_4'\left(\alpha\left(R\right)\right)\,\alpha'\left(R\right).
\end{align*}
{\color{\coloremph}
The hypothesis (H1) is related to the well-known (e.g., \cite{brennen1995}) shape of function $f_1$ (see Fig. \ref{fig:f1}), having a unique critical point $R_\textnormal{crit}$.}

\end{remark}

The next result is a particular case of Theorem 4.2 in \cite{papanicolaou1978}
\begin{proposition}\label{prop:regularity-lions} Let $\Omega$ be a smooth domain on $\mathbb{R}^N$, $f\in \honedual$ and $u\in \hzero$ be the unique solution of the elliptic problem\footnote{Henceforth we denote $\nabla_x \cdot f$ by $\nabla \cdot f$.}
\begin{nalign*}
\nabla\cdot \left(a\,\nabla\,u\right)&=f&&,\\
u&=0 &&\textnormal{ on } \partial \Omega,
\end{nalign*}
for $a\in B_{\alpha,\beta}$, $0<\alpha<\beta$. Then there exists $q>2$ (which depends on $\alpha$, $\beta$, $\Omega$ and on the dimension $N$) such that, if $f\in \sob{-1}{q}$, then $u$ belongs to $\sobz{1}{q}$ and satisfies
$$\left\lVert u \right\rVert_{\sobz{1}{q}}\leq C \norm{f}{-1,q},$$
where $C=C(\alpha,\beta,\Omega,N)$.
\end{proposition}
Now, to fix henceforth a Sobolev space $\sob{1}{q}$, we define the open subset $Q\subset C\left(\bar{\Omega}\right)$ as 
\begin{equation}{\color{\coloremph}
Q=\left\{R\in C\left(\bar{\Omega}\right):R(x)>0~~~
\forall x\in\bar{\Omega}\right\}},\label{eq:def-ensemble-Q}
\end{equation}
and set $q>2$ given by Proposition \ref{prop:regularity-lions} with $\alpha=m_0^3\,m_3\min\{m_1,1\}$ and $\beta=M_0^3\,M_3\max\{M_1,1\}$. 

We define also the mapping
\begin{equation}
\begin{array}{cccc}
A:&Q\times \cont&\longrightarrow& \cont\\
&(R_1,R_2)& \longrightarrow & A_1(R_1)+A_2\left(R_1,R_2\right),
\end{array}
\end{equation}
where $A_1:Q\mapsto \cont$ is such that $A_1\left(R_1\right)$ is the unique solution of the elliptic problem
\begin{nalign}
\nabla\cdot \left(h^3f_3\left(R_1\right)\nabla A_1\left(R_1\right)\right)&=\nabla\cdot \left(\mathbf{U}\,h\,f_4(R_1)\right)&&\textnormal{ in }\Omega,\\
A_1\left(R_1\right)&=0&&\textnormal{ on } \partial \Omega,
\label{eq:def-A1}
\end{nalign}
and
$A_2:Q\times \cont \mapsto \cont$ is such that $A_2\left(R_1,R_2\right)$ is the unique solution of the elliptic problem
\begin{equation}
\begin{aligned}
\nabla\cdot \left(h^3f_3\left(R_1\right)\nabla A_2\left(R_1,R_2\right)\right)&=h\,f_5\left(R_1\right)R_2
&&\textnormal{ in }\Omega,\\
A_2\left(R_1,R_2\right)&=0 &&\textnormal{ on } \partial \Omega.
\label{eq:def-A2}
\end{aligned}
\end{equation}

\begin{remark}\label{remark:cont-A}
Both the solutions of \eqref{eq:def-A1} and \eqref{eq:def-A2} are in $\cont$ since $\sob{1}{p}\subset \cont$ continuously for any $p>N$.
\end{remark}
\begin{remark}\label{remark:A2-linear-R2}
For any $R_1\in Q$, $A_2\left(R_1,\cdot\right)$ is a bounded linear operator.
\end{remark}
\begin{lemma}
	\label{lemma:AC2}
	The application $A$ is of class $C^2$ from $Q\times \cont$ into $\cont$.
	\begin{proof} 
		
		Let us define $\phi:Q\times \cont\times\sobz{1}{q}\mapsto \sob{-1}{q}$ by
		\begin{equation}
		\phi\left(R_1,R_2,p\right)=\nabla\cdot \left(h^3f_3 \left(R_1\right)\nabla\,p\right)-\nabla\cdot \left(\mathbf{U}\,h\,f_4(R_1)\right)-h\,f_5\left(R_1\right)R_2.\label{eq:def:phi}
		\end{equation}
		We show first that $\phi$ is of class $C^2$. Since $f_3$, $f_4$ and $f_5$ are of class $C^2$, it is enough to prove that the application
		$\phi^1:\cont^4 \times \sobz{1}{q}\mapsto \sob{-1}{q}$ defined by
		$$\phi^1\left(\xi_1,\xi_2,\xi_3,\xi_4,w\right)=\nabla\cdot\left(h^3 \xi_1\nabla\,w\right)-\nabla\cdot\left(\mathbf{U}\,h\,\xi_2\right)- h\,\xi_3\, \xi_4$$
		is of class $C^2$, which follows from observing that its first and third terms are quadratic and the second one is linear. 
		
		By the Lax-Milgram Theorem and Proposition \ref{prop:regularity-lions} we have also that the partial derivative
		\begin{equation*}
		\begin{array}{cccc}
		\parder{\phi}{p}\left(R_1,R_2,p\right)\left(z\right):&\sobz{1}{q}&\longrightarrow&  \sob{-1}{q}\\
		&z& \longrightarrow & \nabla\cdot \left(h^3f_3 \left(R_1\right)\nabla\,z\right)
		\end{array}
		\end{equation*} 
		is an isomorphism. Therefore, the result follows from noticing that $\phi\left(R_1,R_2,A\left(R_1,R_2\right)\right)=0$ $\forall\left(R_1,R_2\right)\in Q\times \cont$ and applying the Implicit Function Theorem (e.g., \cite{lang1999}) to the application $\phi$.
	\end{proof}
\end{lemma}
	For a linear operator $\mathcal{L}$ we denote by $\Vp{\mathcal{L}}$ its set eigenvalues and by $\Sp{\mathcal{L}}$ its spectrum. Let us recall the following classical results on Ordinary Differential Equations (ODE) in Banach spaces:
\begin{proposition}
	\label{prop:continuity-spec} Let $X$ be a Banach space, let $A$ be a bounded linear operator on $X$ and $\epsilon>0$. 
	Then there exists $\delta>0$ such that, if $B$ is a bounded linear operator on $X$ and $\norm{A-B}{}<\delta$, then for every $\lambda\in \Sp{B}$ there exists $\xi\in\Sp{A}$ such that $|\lambda-\xi|<\epsilon$.
\end{proposition}
For a detailed proof of the previous result the reader is referred to Lemma 3 in \cite{dunford1957}.
{\color{\coloremph}
\begin{proposition}[Cauchy-Lipschitz]\label{prop:cauchy-lipschitz}
	Let $f\in C(U; E)$, where $U$ is an open set of $E$ and $u_0\in U$, and assume that $f$ is of class $C^r$, $r\in \mathbb{N}^*$. Then the next properties hold
	\begin{itemize}
		\item There exists $T>0$ and $u$ in $C^1([t_0-T,t_0+T];U)$ solution to the Cauchy problem:
		\begin{equation}
		\left\{
		\begin{array}{l}
		u'=f(u)~,\\
		u(t_0)=u_0~.
		\end{array}
		\right.\label{eq:cauchy-problem}
		\end{equation}
		\item If $v$ is another solution of \eqref{eq:cauchy-problem}. Then $v=u$ on the intersection of the intervals of definition of $v$ and $u$.
		\item $u$ is of class $C^{r+1}$.
	\end{itemize}
\end{proposition}

\begin{definition} A solution $u\in C^1([0,T];E)$ of the autonomous Cauchy problem $u'=f(u)$, $u(0)=u_0$ is called \textbf{maximal} if $u$ cannot be extended to a solution on an interval containing $[0,T]$.
\end{definition}

\begin{definition}\label{def:stability}
	Let $f\in C(U;E)$ and $v\in U$. The point $v$ such that $f(v)=0$ is said to be an asymptomatically stable solution for the ODE $u'=f(u)$ if there exist $\epsilon>0$ such that for any $u_0$ such that $\|u_0-v\|\leq \epsilon$, the maximal solution of $u'=f(u)$, $u(0)=u_0$ is well defined for every $t\geq 0$, $\|u(t)-v\|\leq \epsilon$ for every $t\geq 0$ and $\lim_{t\rightarrow \infty}\|u(t)-v\|=0$.
\end{definition}
\begin{definition}\label{def:instability}
	Let $f\in C(U;E)$ and $v\in U$. The point $v$ such that $f(v)=0$ is said to be an unstable stationary solution for the ODE $u'=f(u)$ if there exists $\epsilon_0>0$ such that for every $\eta>0$ there exists $T>0$ and a solution $u\in C^1([0,T],E)$ of $u'=f(u)$ that accomplished $\|u(0)-v\|\leq \eta$ and $\|u(T)-v\|\geq \epsilon_0$.
\end{definition}
\begin{proposition}\label{prop:stability}
	Let $f\in C^2(U;E)$ and $v\in U$ be such that $f(v)=0$. Assume that $\Sp{\textnormal{D}f(v)}\subset\{\lambda\in \mathbb{C}:\textnormal{Re}\,\lambda<0\}$. Then $v$ is an asymptomatically stable solution for the ODE $u'=f(u)$.
\end{proposition}

\begin{proposition}\label{prop:instability}
	Let $f\in C^2(U;E)$ and $v\in U$ be such that $f(v)=0$. Suppose that $\max\{\textnormal{Re}\,\lambda: \lambda\in \Sp{\textnormal{D}f(v)}\}$ is reached at an eigenvalue of $\textnormal{D}f(v)$ with real part strictly positive. Then $v$ is an unstable solution for the ODE $u'=f(u)$.
\end{proposition}

These proofs of propositions \ref{prop:cauchy-lipschitz} to \ref{prop:instability} can be found in \cite{Benzoni2010}, sections 5.4, 8.1 and 8.2.}

\section{Well-posedness with inertial terms}
\label{sec:mathematical-analysis-with-inertial-terms}
Due to the fact that the unknown field $p$ in Eq. \eqref{eq:abstract-rp} can be expressed as an operator depending of $R$ and $\parder{R}{t}$ according to \eqref{eq:abstract-reynolds}, the theory of ODE on Banach spaces can be applied to study the system \eqref{eq:abstract-rp}-\eqref{eq:abstract-cond}.
\subsection{Existence of a local solution}
Let us denote $R_1=R$, $R_2=\parder{R}{t}$ and $\tilde{R}=\begin{pmatrix}
R_1\\ R_2
\end{pmatrix}$. Then, the problem \eqref{eq:abstract-rp}-\eqref{eq:abstract-reynolds}-\eqref{eq:abstract-cond} can be rewritten as
\begin{equation}
	\begin{split}
	\frac{d\tilde{R}}{dt}&=F(\tilde{R}),\\
	\tilde{R}(0)&=\tilde{R}_0,
	\end{split}\label{eq:problem-rewritten}
\end{equation}
where $\tilde{R}_0=\begin{pmatrix}
{r}_1\\ {r}_2
\end{pmatrix}\in Q\times\cont$ and $F:Q\times \cont\mapsto \left(\cont\right)^2$ with
\begin{equation}
F(R_1,R_2)=\begin{pmatrix}
R_2\\-\frac{3}{2}\frac{R_2^2}{R_1}-R_2f_2\left(R_1\right)+\frac{f_1\left(R_1\right)-A\left(R_1,R_2\right)}{R_1}
\end{pmatrix}.\label{eq:def:F(R1,R2)}
\end{equation}
By means of Lemma \ref{lemma:AC2} we have that $F$ is of class $C^2$. Thus, from the Cauchy-Lipschitz Theorem we obtain the next  local existence and uniqueness result:
\begin{theorem}There exists $T>0$ such that the  problem \eqref{eq:problem-rewritten} has a unique solution in $C^3\left([0,T];Q\times\cont\right)$.
\end{theorem}
\subsection{Existence of stationary solutions}
\label{sec:existence-stationary}

Observe that a stationary solution $(R_s,p_s)$ of problem \eqref{eq:abstract-rp}-\eqref{eq:abstract-reynolds} satisfies
$p_s=f_1\left(R_s\right)$. For the next result we denote $h^+=h-h_0$ (notice that $h^+=0$ if and only if $h$ is constant). Thus $(R_s,p_s)$ is solution of the system
\begin{nalign}
\nabla\cdot \left(\left(h^++h_0\right)^3f_3\left(R_s\right)\nabla p_s\right)&=\nabla\cdot\left(\mathbf{U}\left(h^++h_0\right)f_4\left(R_s\right)\right)&&\textnormal{ in } \Omega,\\
		p_s&=f_1\left(R_s\right)&&\textnormal{ in } \Omega,\\
	p_s&=0&&\textnormal{ on }\partial \Omega.
\label{eq:problem-stationary}
\end{nalign}
Notice that in the particular case $h^+=0$ or $\mathbf{U}=0$, $\left(R_s,p_s\right)=\left(\bar{R},0\right)$ is solution of \eqref{eq:problem-stationary}, with $\bar{R}$ given in (H1).
\begin{theorem}
	\label{theo:existence_U_fix}
	Fix $\mathbf{U}\in\mathbb{R}^2$ and $h_0>0$. Then the problem \eqref{eq:problem-stationary} has a unique solution $\left(R_s,p_s\right)$ with $R_s>0$ whenever $\norm{h^+}{\infty}$ is small enough. Moreover, the solution $\left(R_s,p_s\right)$ depends continuously on $h^+$.
	\begin{proof}
First we use the relation $p_s=f_1\left(R_s\right)$ to rewrite the stationary problem. Since $\nabla p_s=f_1'\left(R_s\right)\nabla R_s$, making the change of variable $R_s=\bar{R}+\xi$  problem \eqref{eq:problem-stationary} can be written in function of $\xi$ as
\begin{nalign}
-\nabla\cdot \left(\left(h^++h_0\right)^3a_0\left(\xi\right)\nabla\,\xi\right)&=\nabla\cdot \left(\mathbf{U} h^+\,b_0(\xi)\right)+\nabla\cdot \left(\mathbf{U}h_0\,b_0(\xi)\right)&&\textnormal{ in }\Omega,\\
\xi&>-\bar{R}&&\textnormal{ in }\Omega,\\
\xi&=0&&\textnormal{ on } \partial \Omega,
\label{eq:problem-stationary-change}
\end{nalign}
where $a_0\left(\xi\right)=-f_3\left(\bar{R}+\xi\right)f_1'\left(\bar{R}+\xi\right)$, and $b_0(\xi)=f_4\left(\bar{R}+\xi\right)$. We introduce the set
$$W={\color{\coloremph}\left\{\xi\in W_0^{1,q}\left(\Omega\right):\underset{\Omega}{\essinf}~\xi>-\bar{R}\right\}},$$
which is open since the continuous embedding $W_0^{1,q}\subset C\left(\bar{\Omega}\right)$, and the application
\begin{equation}
\begin{array}{cccl}
\phi_2:&W\times L^\infty\left(\Omega\right)&\longmapsto& W^{-1,q}\left(\Omega\right)\\
&(\xi,\delta)& \longmapsto & \nabla\cdot \left(\left(\delta+h_0\right)^3a_0\left(\xi\right)\nabla\,\xi\right)+\nabla\cdot \left(\mathbf{U}\delta\,b_0(\xi)\right)+\nabla\cdot \left(\mathbf{U}h_0\,b_0(\xi)\right).
\end{array}
\end{equation}
Using an argument analogous to the one used in Lemma \ref{lemma:AC2} to prove that $\phi^1$ is of class $C^2$, it is possible to prove that $\phi_2$ is of class $C^2$. Now noticing that $\phi_2\left(0,0\right)=0$, let us assume that $\parder{\phi_2}{\xi}\left(0,0\right)$ is invertible. Then, by means of the Implicit Function Theorem, we have that 
\begin{itemize}
	\item $\exists V_1\subset W$ neighborhood of $0$ on $W_0^{1,q}\left(\Omega\right)$; $V_2$ neighborhood of $0$ on $L^\infty\left(\Omega\right)$;
	\item $\exists \psi:V_2\longmapsto V_1$ function of class $C^1$ such that $\forall \delta \in V_2$, $\psi(\delta)$ is solution of problem \eqref{eq:problem-stationary-change}. Equivalently, $\left(R_s,p_s\right)=\left(\bar{R}+\psi(\delta),f_1\left(\bar{R}+\psi(\delta)\right)\right)$ is solution of problem \eqref{eq:problem-stationary}.
\end{itemize}
which is the result we want as the existence of $V_2$ can also be described as $\norm{h^+}{\infty}$ small enough.

It only remains to show that $\parder{\phi_2}{\xi}\left(0,0\right)$ is invertible. Indeed, we have $\forall z \in\sobz{1}{q}$:
\begin{align*}
\parder{\phi_2}{\xi}\left(0,0\right)\left(z\right)&=\nabla\cdot\left.\left(\left(h_0+\delta\right)^3\left(a_0\left(\xi\right) \nabla z+a_0'\left(\xi\right) z\, \nabla \xi \right)+\left(h_0+\delta\right)\,b_0'\left(\xi\right)\mathbf{U}z\right)\right|_{\left(\xi,\delta\right)=\left(0,0\right)}\\
&=\nabla\cdot\left(h_0^3\,a_0\left(0\right) \nabla z+h_0\,b_0'\left(0\right)\mathbf{U}z\right).
\end{align*}
 Fixing an arbitrary $g\in \sob{-1}{q}$ and denoting $\ell=h_0b'_0\left(0\right)\mathbf{U} \in \mathbb{R}^2$ we will prove that there exists a unique $z\in \sobz{1}{q}$ such that 
$$\nabla\cdot\left(h_0^3\,a_0\left(0\right)\nabla z+\ell\,z\right)=g\qquad\mbox{in }\Omega.$$
Since $g\in H^{-1}\left(\Omega\right)$, $a_0(0)>0$ and $h_0>0$ (see (H1) and (H3)), by means of the Lax-Milgram Theorem the variational problem
$$-\int_\Omega \left(h_0^3\,a_0\left(0\right)\nabla z+\ell\, z\right)\cdot \nabla \phi\, d\Omega=\int_\Omega g\,\phi\,d\Omega\qquad \forall \phi\in \hzero,$$
has a unique solution $z\in \hzero$. Moreover, from the continuous inclusion $H^1\left(\Omega\right)\subset L^q\left(\Omega\right)$, we have $\nabla\cdot\left(\ell\, z\right)\in\sob{-1}{q}$ and thus by Proposition \ref{prop:regularity-lions} we obtain $z\in \sobz{1}{q}$.
\end{proof} 
\end{theorem}

A proof analogous to the one of Theorem \ref{theo:existence_U_fix} may be written for the next result:
\begin{theorem}\label{theo:existence_h_fix}
	Fix $h\in B_{m_0,M_0}$, $0<m_0<M_0$. Then there exists $\epsilon\left(h\right)>0$ such that the problem \eqref{eq:problem-stationary} has a unique solution $\left(R_s,p_s\right)$ with $R_s>0$ whenever $\norm{\mathbf{U}}{}<\epsilon\left(h\right)$.  Moreover, the solution $\left(R_s,p_s\right)$ depends continuously on $\mathbf{U}$.
\end{theorem}

\subsection{Stability Analysis}
\label{sec:stability-withinertia}

Recalling the application $F$ given by \eqref{eq:def:F(R1,R2)} and the stationary solution $(R_s,p_s)$ introduced in the previous section, we denote by $\mathcal{L}_F$ the differential of $F$ at $\left(R_s,0\right)$, i.e.,
\begin{equation}
\begin{array}{cccl}
\mathcal{L}_F:&\left(\cont\right)^2&\longmapsto& \left(\cont\right)^2\\
&(S_1,S_2)& \longmapsto & D\,F\left(R_s,0\right)\left(S_1,S_2\right).\label{eq:jacobian_inertia}
\end{array}
\end{equation} 
We will show the stability of the stationary solution in some particular cases. For this, we will show that the spectrum of $\mathcal{L}_F$ is such that $\mbox{Re}\left(\lambda\right)<0~\forall \lambda\in \Sp{\mathcal{L}_F}\setminus\{0\}$. Previously, we perform some computations.

Recalling that $f_1\left(R_s\right)=p_s=A\left(R_s,0\right)$ we obtain:
\begin{align}
\left(\mathcal{L}_F\left(S_1,S_2\right)\right)_1&=S_2,\\
\left(\mathcal{L}_F\left(S_1,S_2\right)\right)_2&=\frac{f_1'\left(R_s\right)S_1}{R_s}-\frac{1}{R_s}\left(D_1A\left(R_s,0\right)\left(S_1\right)+D_2A\left(R_s,0\right)\left(S_2\right)\right)+\frac{1}{R_s^2}A\left(R_s,0\right)S_1-f_2\left(R_s\right)S_2.
\end{align}
Now, since $A_2\left(R,0\right)=0$ for any $R$ in $Q$, we have that $D_1 A\left(R_s,0\right)=D A_1\left(R_s\right)$. With this, deriving \eqref{eq:def-A1} with respect to $R_1$ and denoting $\pi_1\left(S_1\right)=D_1A\left(R_s,0\right)\left(S_1\right)$ we obtain that $\pi_1\left(S_1\right)$ satisfies
\begin{align}
\begin{split}
-\nabla\cdot \left(h^3f_3\left(R_s\right) \,\nabla \pi_1\left(S_1\right)\right)&=\nabla\cdot \left(h^3f_3'\left(R_s\right) S_1 \,\nabla A_1\left(R_s\right)-\mathbf{U}hf_4'\left(R_s\right)S_1\right),\\
\pi_1\left(S_1\right)&=0 \qquad\textnormal{on } \partial \Omega.
\end{split}
\label{eq:def:pi1}
\end{align}
Similarly, we have $D_2 A\left(R_s,0\right)\left(S_2\right)=D_2 A_2\left(R_s,0\right)\left(S_2\right)=A_2\left(R_s,S_2\right)$. Thus, denoting $\pi_2\left(S_2\right)=D_2A\left(R_s,0\right)\left(S_2\right)$ we have that $\pi_2\left(S_2\right)$ accomplishes
\begin{nalign}
-\nabla\cdot \left(h^3f_3\left(R_s\right) \,\nabla \pi_2\left(S_2\right)\right)&=-hf_5\left(R_s\right)S_2&&\textnormal{ in }  \Omega,\\
\pi_2\left(S_2\right)&=0 &&\textnormal{ on } \partial \Omega.
\label{eq:def:pi2}
\end{nalign}
For the next results we denote  $b_1=-f_1'(\bar{R})\bar{R}^{-1}>0$, $b_2=f_2\left(\bar{R}\right)>0$, $b_r=1/\bar{R}$, $b_3=f_3\left(\bar{R}\right)$, $b_4=-f_4'\left(\bar{R}\right)$ and $b_5=-f_5\left(\bar{R}\right)$, all positive constants as follows from (H1)-(H5).
\begin{remark}\label{remark:trivial-case}
 If $h^+=0$ or $\mathbf{U}=0$ we have $A_1\left(R_s\right)=0$. Thus $A(R_s,0)=p_s=0$, $R_s=\bar{R}$ and $\mathcal{L}_F\left(S_1,S_2\right)$ can be written
\begin{equation}\label{eq:L_F}
	\mathcal{L}_F\left(S_1,S_2\right)=B\begin{pmatrix}
	S_1\\S_2
	\end{pmatrix}-b_r\begin{pmatrix}
	0\\\pi_1\left(S_1\right)+\pi_2\left(S_2\right)
	\end{pmatrix},
\end{equation}
where $B=\begin{pmatrix}0&1\\-b_1&-b_2
\end{pmatrix}$, and Eq. \eqref{eq:def:pi1} reads
\begin{nalign}
-b_3\nabla\cdot \left(h^3 \,\nabla \pi_1\left(S_1\right)\right)&=b_4\nabla\cdot \left(\mathbf{U}hS_1\right)&&\textnormal{ in }  \Omega,\\
\pi_1\left(S_1\right)&=0 &&\textnormal{ on } \partial \Omega.
\label{eq:pi1-triv-case}
\end{nalign}
We denote by $\{\lambda_1^B,\lambda_2^B\}$ the set of eigenvalues of $B$ and notice that $\mbox{Re}\left(\lambda_1^B\right)<0$ and $\mbox{Re}\left(\lambda_2^B\right)<0$.
\end{remark}
\begin{lemma}\label{lemma:spec-L_F}
	Let $h^+=0$ or $\mathbf{U}=0$. Then
	\begin{equation*}
\Sp{\mathcal{L}_F}\subset \Vp{\mathcal{L}_F}\cup \{\lambda_1^B,\lambda_2^B\}.
	\end{equation*}
	Moreover if $\lambda\in \Vp{\mathcal{L}_F}\setminus\{\lambda_1^B,\lambda_2^B\}$ with associated eigenfunction $\left(S_1,S_2\right)\in\cont^2$ then $\left(S_1,S_2\right)\in\hzero^2$, $S_2=\lambda S_1$ and $S_1$ is solution of the problem
	\begin{align}
	\frac{b_3}{b_r}\xi\left(\lambda\right)\nabla\cdot\left(h^3\nabla S_1\right)&=b_4 \mathbf{U}\cdot\nabla\left(hS_1\right)+\lambda\, b_5 h\, S_1\label{eq:eigenfunction-L_G1} && \textnormal{in }\Omega,\\
	S_1&=0&&\textnormal{on }\partial\Omega,\label{eq:eigenfunction-L_G2}
	\end{align}
	where $\xi\left(\lambda\right)=\lambda^2+b_2\lambda+b_1$ with roots $\{\lambda_1^B,\lambda_2^B\}$.
\begin{proof}
	Remind that $p_s=A(R_s,0)=0$ and $R_s=\bar{R}$. For any $\lambda\in \mathbb{C}\setminus \{\lambda_1^B,\lambda_2^B\}$, from Eq. \eqref{eq:L_F} we have
	\begin{equation*}
	\left(\mathcal{L}_F-\lambda I\right)\begin{pmatrix}
	S_1\\S_2
	\end{pmatrix}=\left(B-\lambda I\right)\left[\begin{pmatrix}
	S_1\\S_2
	\end{pmatrix}-b_r\left(B-\lambda I\right)^{-1}
	\begin{pmatrix}
	0\\ \pi_1\left(S_1\right)+\pi_2\left(S_2\right)
	\end{pmatrix}	
	\right].\label{eq:vp-L}
	\end{equation*}	
	Since the map $\left(S_1,S_2\right)\mapsto \pi_1\left(S_1\right)+ \pi_2\left(S_2\right)$ is compact, by means of the Fredholm's Alternative Theorem the mapping at the right hand side of this equation (from $\cont^2$ into itself) is injective if and only if it is surjective, from where we have the $\Sp{\mathcal{L}_F}\subset \Vp{\mathcal{L}_F}\cup \{\lambda_1^B,\lambda_2^B\}$.
	
	Fix now $\lambda\in \Vp{\mathcal{L}_F}\setminus\{\lambda_1^B,\lambda_2^B\}$ with associated eigenvector $\left(S_1,S_2\right)\neq\left(0,0\right)$ so we can write
	\begin{align*}
	S_2=\lambda S_1,\\
	-b_1S_1-b_2S_2-b_r\left[\pi_1\left(S_1\right)+\pi_2\left(S_2\right)\right]=\lambda S_2.
	\end{align*}
	Then we obtain
	$$\pi_2\left(\lambda S_1\right)+\pi_1\left(S_1\right)=-\frac{\xi\left(\lambda\right)}{b_r}S_1.$$
	Since $\xi\left(\lambda\right)\neq 0$ and from the definitions of $\pi_1$ and $\pi_2$ we deduce that $\left(S_1,S_2\right)\in\hzero^2$. Thus, using this last equation, Eq. \eqref{eq:def:pi2} and Eq. \eqref{eq:pi1-triv-case} we obtain the Eqs. \eqref{eq:eigenfunction-L_G1}-\eqref{eq:eigenfunction-L_G2}.
\end{proof}	
\end{lemma}

\begin{theorem}\label{theo:stabilite-avec-inertie-h-fixe} Let $h$ be as in Theorem \ref{theo:existence_h_fix}. Then there exists $\epsilon=\epsilon\left(h\right)>0$ such that if $\norm{\mathbf{U}}{\infty}<\epsilon$ the solution $\left(R_s,p_s\right)$ of problem \eqref{eq:problem-stationary} is asymptotically stable for the evolution problem \eqref{eq:problem-rewritten}.
\begin{proof} Assume first $\mathbf{U}=0$ and denote $\mathcal{L}_F^0=\left.\mathcal{L}_F\right|_{\mathbf{U}=0}$. Then due to Lemma \ref{lemma:spec-L_F} it is enough to study the eigenvalues of $\mathcal{L}_F$. Thus, take $\lambda\in \Vp{\mathcal{L}_F}\setminus\{\lambda_1^B,\lambda_2^B\}$ with associated eigenfunction $\left(S_1,S_2\right)$, from Lemma \ref{lemma:spec-L_F} we have $S_2=\lambda S_1$ and $S_1\in \hzero$ accomplishing Eqs. \eqref{eq:eigenfunction-L_G1} and \eqref{eq:eigenfunction-L_G2}, which read 
\begin{nalign*}
\frac{b_3}{b_r}\xi\left(\lambda\right)\nabla\left(h^3\nabla S_1\right)&=\lambda\, b_5 h\, S_1&& \textnormal{ in }\Omega,\\
S_1&=0&&\textnormal{ on }\partial\Omega.
\end{nalign*}
Since $\xi\left(\lambda\right)$ is not null we deduce that $\lambda\neq 0$, otherwise $\left(S_1,S_2\right)$ would be null. Then we obtain that $S_1$ accomplishes the next variational formulation
\begin{equation}
-\frac{b_3}{b_r}\frac{\xi\left(\lambda\right)}{\lambda}\int_\Omega h^3 \nabla S_1\nabla \phi\,d\Omega=b_5\int_\Omega h S_1\,\phi\,d\Omega\qquad \forall \phi \in \hzero.
\end{equation}
Taking $\phi=S_1$ we obtain that $\gamma=-\xi\left(\lambda\right)/\lambda\in \mathbb{R}^+$ and since $\lambda$ accomplishes the equation $\lambda^2+\left(\gamma +b_2\right)\lambda+b_1=0$ we conclude that $\mbox{Re}\left(\lambda\right)<0$. We have shown the result for the case $\mathbf{U}=0$.

For the general case, we observe from Theorem \ref{theo:existence_h_fix} that the mapping $\mathbf{U}\mapsto R_s\left(\mathbf{U}\right)$ is continuous in a neighborhood $V_1\ni 0$ in $\mathbb{R}^2$, thus if $\mathbf{U}\rightarrow 0$ in $\mathbb{R}^2$ then $\norm{DF\left(R_s\left(\mathbf{U}\right),0\right)-DF\left(\bar{R},0\right)}{}\rightarrow 0$ in the space of linear continuous operators from $\cont^2$ into itself. Then the result follows from Proposition \ref{prop:continuity-spec}.
\end{proof}
\end{theorem}
We give now a result of instability for $\norm{\mathbf{U}}{}$ big enough.
\begin{theorem}\label{theo:instabilite-avec-inertie-U-fixe}{\color{\coloremph} Let us assume $h^+=0$ and $\Omega=\,]0,1[\times ]0,1[$. Then the solution $\left(R_s,p_s\right)$ of problem \eqref{eq:problem-stationary} is asymptotically unstable for the evolution problem \eqref{eq:problem-rewritten} for $\|\mathbf{U}\|$ big enough.}
\begin{proof}
{\color{\coloremph}
 Due to Lemma \ref{lemma:spec-L_F} it is enough to study the eigenvalues of $\mathcal{L}_F$. Fix now $\lambda\in \Vp{\mathcal{L}_F}\setminus\{\lambda_1^B,\lambda_2^B\}$ with associated eigenvector $\left(S_1,S_2\right)\neq\left(0,0\right)$. Now defining $\gamma_1,\gamma_2\in \mathbb{C}$ by
$$\gamma_1=-\frac{b_4\,b_r}{h_0^2b_3\xi\left(\lambda\right)},\qquad \gamma_2=-\frac{b_5\,b_r}{h_0^2b_3\xi\left(\lambda\right)},$$
then from Eqs. \eqref{eq:eigenfunction-L_G1}-\eqref{eq:eigenfunction-L_G2} we have
\begin{align*}
\Delta S_1+\gamma_1 \mathbf{U}\cdot\nabla\left(S_1\right)+\lambda\gamma_2\, S_1&=0 && \textnormal{in }\Omega,\\
S_1&=0&&\textnormal{on }\partial\Omega.
\end{align*}
We deduce from this that $\lambda\neq 0$. In fact, if $\lambda=0$ then one may compute that $S_1=0,S_2=\lambda S_1=0$, which is a contradiction. Assuming $S_1(x_1,x_2)=\varphi_1(x_1)\,\varphi_2(x_2)$ with both $\varphi_1$ and $\varphi_2$ non nulls and $\varphi_1(0)=\varphi_1(1)=\varphi_2(0)=\varphi_2(1)=0$ it is possible to obtain
$$\frac{\varphi_1''(x_1)}{\varphi_1(x_1)}+\gamma_1U_1 \frac{\varphi_1'(x_1)}{\varphi_1(x_1)}=-\frac{\varphi_2''(x_1)}{\varphi_2(x_2)}-\gamma_1U_2 \frac{\varphi_2'(x_2)}{\varphi_2(x_2)}-\lambda\gamma_2.$$
Therefore there exists $\mu\in \mathbb{C}$ such that
\begin{align}
&\varphi_1''(x_1)+\gamma_1U_1\,\varphi_1'(x_1)-\mu\,\varphi_1(x_1)=0 ,\label{eq:phi1}\\
&\varphi_2''(x_2)+\gamma_1U_2\,\varphi_2'(x_2)+(\lambda\gamma_2+\mu)\varphi_2(x_2)=0\label{eq:phi2}.
\end{align}
Denote by $r_1$, $r_2$ the roots of the characteristic polynomial $P\left(r\right)=r^2+\gamma_1U_2r+\lambda\gamma_2+\mu$ of the last equation. Then $r_1\neq r_2$, otherwise $\varphi_2$ would be null, and so $\varphi_2$ can be written
	$$\varphi_2\left(x_2\right)=C_1\exp\left(r_1x_2\right)+C_2\exp\left(r_2x_2\right).$$
 Thus the conditions $\varphi_2(0)=\varphi_2(1)=0$ imply
 \begin{align*}
	 	C_1+C_2&=0,\\
	 	C_1\exp{r_1}+C_2\exp {r_2}&=0.
 \end{align*}
 Thus, since $\left(C_1,C_2\right)\neq \left(0,0\right)$ we have
 $$\mbox{det}\begin{pmatrix}
 1&1\\\exp{r_1} & \exp{r_2}
 \end{pmatrix}=0,$$
 hence $r_1$ and $r_2$ satisfy the equation $r_2-r_1=2k_2\pi \,i$ $\forall k_2\in \mathbb{N}^*$, from which we deduce that
 \begin{equation}
 \gamma_1^2\,U_2^2-4(\lambda\gamma_2+\mu)=-4\,k_2^2\pi^2,\qquad\forall k_2\in \mathbb{N}^*,\label{eq:k2}
 \end{equation}
 where we have used the fact that $r_1+r_2=-\gamma_1 U_2$ and $r_1r_2=\lambda\gamma_2+\mu$. Analogously, from the characteristic polynomial of Eq. \eqref{eq:phi1} one may obtain
 \begin{equation}
 \gamma_1^2\,U_1^2-4(-\mu)=-4\,k_1^2\pi^2,\qquad\forall k_1\in \mathbb{N}^*.\label{eq:k1}
 \end{equation}
Denoting $k=(k_1,k_2)\in\mathbb{N}^*\times\mathbb{N}^*$, the addition of these two equations implies
 $$\gamma_1^2\,\|\mathbf{U}\|^2-4\,\lambda\gamma_2=-4\,\|k\|^2\pi^2.$$
Recalling the definitions of $\gamma_1$ and $\gamma_2$ one concludes that $\lambda$ is root of the fourth degree polynomial given by
	\begin{equation*}
		\begin{split}
		P_k\left(\lambda\right)=4\|k\|^2\pi^2\lambda^4+\left(4\sigma_2+8\pi^2\|k\|^2b_2\right)\lambda^3+\left(4\sigma_2b_2+4\pi^2\|k\|^2\left(b_2^2+2b_1\right)\right)\lambda^2+\\+\left(4\sigma_2b_1+8\pi^2\|k\|^2b_1b_2\right)\lambda+4\pi^2\|k\|^2b_1^2+\sigma_1 \|\mathbf{U}\|^2,
	\end{split}
	\end{equation*}
	 where $\sigma_1=\frac{b_4^2b_r^2}{b_3^2h_0^4}$ and $\sigma_2=\frac{b_5b_r }{b_3h_0^2}$ are both positive constants. Rewriting this polynomial as
	 $P_k\left(\lambda\right)=\alpha_0\lambda^4+\beta_0\lambda^3+\alpha_1\lambda^2+\beta_1\lambda+\alpha_2$, let us now denote the Hurwitz determinants associated to $P_k$:
 $$\Delta_1=\mbox{det}\begin{pmatrix}
 \beta_0 \\
 \end{pmatrix},~
\Delta_2=\mbox{det}\begin{pmatrix}
 \beta_0 & \beta_1\\ \alpha_0 & \alpha_1 \\
 \end{pmatrix},~\Delta_3=\mbox{det}\begin{pmatrix}
 \beta_0 & \beta_1 & 0\\ \alpha_0 & \alpha_1 & \alpha_2 \\
 0 & \beta_0& \beta_1 
 \end{pmatrix},~\Delta_4=\mbox{det}\begin{pmatrix}
 \beta_0 & \beta_1 & 0 & 0\\ \alpha_0 & \alpha_1 & \alpha_2 & 0 \\
 0 & \beta_0& \beta_1 &0\\
  0 & \alpha_0& \alpha_1 &\alpha_2
 \end{pmatrix}.$$
Then one obtains $\Delta_1=4\sigma_2+8\pi^2\|k\|^2b_2$, $\Delta_2=\left(4\sigma_2+8\pi^2\|k\|^2b_2\right)\left(4\sigma_2b_2+4\pi^2\|k\|^2\left(b_2^2+b_1\right)\right)$,
 \begin{equation*}
\begin{split}
\Delta_3=\left(320\,b_1b_2^2\pi^2\|k\|^2-16\sigma_1\|\mathbf{U}\|^2\right)\sigma_2^2+\left(512b_1b_2^3\pi^4\|k\|^4-64b_2\pi^2\sigma_1\|k\|^2\|\mathbf{U}\|^2\right)\sigma_2-\\-64b_2^2\pi^4\sigma_1\|k\|^4\|\mathbf{U}\|^2
+256\,b_1b_2^4\pi^6\|k\|^6+64b_1b_2\sigma_2^3,
\end{split}
 \end{equation*}
 and $\Delta_4=\alpha_2 \Delta_3$. According to the Routh-Hurwitz Theorem \cite{Gantmacher1959} the number of roots of the polynomial $P_k$ with positive real part is equal to the total number of changes of sign in the sequence $\{\alpha_0,\Delta_1,\frac{\Delta_2}{\Delta_1},\frac{\Delta_3}{\Delta_2},\frac{\Delta_4}{\Delta_3}\}$. One may compute $\alpha_0>0$, $\Delta_1>0$, $\frac{\Delta_2}{\Delta_1}>0$, $\frac{\Delta_4}{\Delta_3}>0$ and
$\frac{\Delta_3}{\Delta_2}<0$ for $\|\mathbf{U}\|$ big enough, which ends the proof by using Proposition \ref{prop:instability}. }
\end{proof}	
\end{theorem}
		
\section{Well-posedness without inertial terms}
\label{sec:mathematical-analysis-without-inertial-terms}

Disregarding the inertial terms in Eq. \eqref{eq:abstract-rp} (as done in \cite{natsumeda1987,snyder2016,braun2017}) we obtain the following simplified version of the Rayleigh-Plesset equation
\begin{equation}
\parder{R}{t}=\frac{f_1\left(R\right)-p}{Rf_2\left(R\right)}\label{eq:abstract-rp-inertialess},
\end{equation}
along the initial condition
\begin{equation}
R\left(x,0\right)=r_1\left(x\right)\qquad \forall x\in \Omega,\label{eq:bc-rp-inertialess}
\end{equation}
where $r_1\in\cont$ known and $p\in \sobz{1}{q}$ is the solution of \eqref{eq:abstract-reynolds}.

\subsection{Existence of a local solution}
Let us prove that we can express $\partial R/ \partial t$ as a function of $R$ from \eqref{eq:abstract-rp-inertialess}. Denoting $R_1=R$, $R_2=\parder{R}{t}$, we recall the decomposition
$$p=A\left(R_1,R_2\right)=A_1(R_1)+A_2(R_1,R_2),$$
with $A_1$ and $A_2$ as in \eqref{eq:def-A1} and \eqref{eq:def-A2} respectively.
 Now, defining $\Pi:Q\times \cont\mapsto \cont$ by
\begin{equation}
\Pi\left(R_1,R_2\right)=\frac{f_1\left(R_1\right)-A_1(R_1)-A_2(R_1,R_2)}{R_1f_2\left(R_1\right)},\label{eq:def-PI}
\end{equation}
we have the next result:
\begin{lemma}\label{lemma:fixed-point-Pi}
	Given $R\in Q$, there exists a unique $G\left(R\right)\in \cont$ such that
	$$G\left(R\right)=\Pi\left(R,G\left(R\right)\right),$$
	and the mapping $R\mapsto G\left(R\right)$ is of class $C^2$.
	\begin{proof}
	Let us fix $R\in Q$, we will show that there exists a unique $S\in\cont$ such that $S=\Pi\left(R,S\right)$. Using \eqref{eq:def-PI}, we first notice that the equation $S=\Pi\left(R,S\right)$ is equivalent to
$$S+\frac{A_2\left(R,S\right)}{Rf_2\left(R\right)}=\frac{f_1\left(R\right)-A_1\left(R\right)}{Rf_2\left(R\right)}.$$
We denote by $J:\cont\mapsto \cont$ the linear mapping $S\mapsto S+A_2\left(R,S\right)/ \left(Rf_2\left(R\right)\right)$. To prove the existence of a unique solution for the last equation we will show that $J$ is bijective, which will give us the existence of $G$ by taking $G\left(R\right)=S$. Now, since the mapping $S\mapsto A_2\left(R,S\right)$ is compact, by means of the Fredholm Alternative Theorem it is enough to prove that $J$ is injective. Indeed, let us take $w\in\cont$ such that $J\left(w\right)=0$, then we have
$$R f_2\left(R\right) w+A_2\left(R,w\right)=0.$$
Multiplying this equation by $-f_5\left(R\right)hw$ and integrating by parts we obtain
\begin{equation}
\int_\Omega Rf_2\left(R\right)\left(-f_5\left(R\right)\right)hw^2\,d\Omega+\int_\Omega \left(-f_5\left(R\right)\right)hA_2\left(R,w\right)\,w\,d\Omega=0.\label{eq:auxiliar-lemma-existence-G}
\end{equation}
		Now, multiplying  \eqref{eq:def-A2} by $A_2\left(R_1,R_2\right)$, integrating and using (H5) we have for any $\left(R_1,R_2\right)\in Q\times \cont$
		\begin{equation}
		\int_\Omega \left(-f_5\left(R_1\right)\right)h\,A_2(R_1,R_2)\,R_2\,d\Omega\geq 0\label{eq:ineq-monotonie}.
		\end{equation}
Taking $R_1=R$ and $R_2=w$ in the last equation and carrying that into Eq. \eqref{eq:auxiliar-lemma-existence-G} we obtain $w=0$, so we conclude $J$ is injective.

Next, we prove that $G$ is of class $C^2$. Let us define the mapping $\Phi:Q\times \cont\mapsto\cont$ such that
		$$\Phi\left(R,S\right)=S-\Pi\left(R,S\right),$$
		which is of class $C^2$ since all the involved functions are regular enough. Now, fixing some arbitrary $\left(R_0,S_0\right)\in Q\times \cont$ such that $\Phi\left(R_0,S_0\right)=0$ we have for any $w\in \cont$
		$$\parder{\Phi}{S}\left(R_0,S_0\right)\left(w\right)=w+\frac{A_2\left(R_0,w\right)}{R_0f_2\left(R_0\right)}=J\left(w\right).$$
		From where we obtain that $\parder{\Phi}{S}\left(R_0,S_0\right)$ is an automorphism on $\cont$. Thus, we conclude that $G$ is of class $C^2$ by means of the Implicit Function theorem.
	\end{proof}
\end{lemma}
\begin{theorem}There exists $T>0$ such that  problem \eqref{eq:abstract-reynolds}-\eqref{eq:abstract-rp-inertialess}-\eqref{eq:bc-rp-inertialess} has a unique solution in $C^3\left([0,T];Q\right)$.
	\begin{proof}
		The result follows directly from applying the Cauchy-Lipschitz Theorem to the equivalent evolution problem\begin{equation}
		\parder{R}{t}=G(R),\label{eq:ev-problem-G}
		\end{equation}
		along the initial condition \eqref{eq:bc-rp-inertialess}.
	\end{proof}
\end{theorem}

\subsection{Stability analysis} 
\label{sec:stability-wo-inertia}
Let us notice the stationary solution of \eqref{eq:abstract-rp-inertialess} is also the couple $\left(R_s,p_s\right)$ obtained in Section \ref{sec:existence-stationary}. Here we study the stability of that solution for the evolution problem \eqref{eq:ev-problem-G}.

Here we denote the derivative
\begin{equation}
\begin{array}{cccl}
\mathcal{L}_G:&\cont&\longmapsto& \cont\\
&w& \longmapsto & D\,G\left(R_s\right)\left(w\right).\label{eq:def-L_G}
\end{array}
\end{equation} 
Using the definition of $\Pi\left(R,S\right)$ we compute the derivative with respect to $R$ in the equation $S=\Pi\left(R,S\right)$ and make the evaluation at $R=R_s$, $S=0$, so we obtain that $\mathcal{L}_G\left(w\right)$ satisfies:
\begin{equation}
R_sf_2\left(R_s\right)\mathcal{L}_G\left(w\right)-f_1'\left(R_s\right)w+\pi_1\left(w\right)+\pi_2\left(\mathcal{L}_G\left(w\right)\right)=0
\label{eq:equation-L_G},
\end{equation}
with $\pi_1$ and $\pi_2$ as in Eqs. \eqref{eq:def:pi1} and \eqref{eq:def:pi2} respectively. 

For the next results we denote
$d_1=-f_1'\left(\bar{R}\right)/\left(\bar{R}f_2\left(\bar{R}\right)\right)$, $d_2=\left(\bar{R}f_2\left(\bar{R}\right)\right)^{-1}$, $d_3=\bar{R}\,f_3\left(\bar{R}\right)f_2\left(\bar{R}\right)$, $d_4=-f_4'\left(\bar{R}\right)$ and $d_5=-f_5\left(\bar{R}\right)$. All these constants are positive as follows from (H1)-(H5).
\begin{lemma}\label{lemma:spec-L_G}
Asumme $h^+=0$ or $\mathbf{U}=0$. Then 
$$\Sp{\mathcal{L}_G}\subset\Vp{\mathcal{L}_G}\cup\{-d_1\}.$$
Moreover, if $w\in\cont$ is an eigenvector of $\mathcal{L}_G$ with associated eigenvalue $\lambda$, then $w\in\hzero$ and it satisfies 
\begin{nalign}
	\begin{split}
	d_3\left(d_1+\lambda\right)\nabla\cdot \left(h^3\nabla w\right)&=d_4\,\mathbf{U}\cdot \nabla \left(hw\right)+\lambda \,d_5\,hw&&\textnormal{ in } \partial \Omega,\\
	w&=0 &&\textnormal{ on } \partial \Omega.
	\end{split}
	\label{eq:eigenvectors-L_G}
\end{nalign}
\begin{proof}
	 From Remark \ref{remark:trivial-case} we have $\left(R_s,p_s\right)=\left(\bar{R},0\right)$. Putting this into Eq. \eqref{eq:equation-L_G}
	we obtain that for any $\lambda\in\mathbb{C}$:
	\begin{equation}
	\mathcal{L}_G\left(w\right)-\lambda\, w=\left(\lambda+d_1\right)\left[-w-\frac{d_2}{\lambda+d_1}\left[\pi_1\left(w\right)+\pi_2\left(\mathcal{L}_G\left(w\right)\right)\right]\right],\label{eq:L_G2}
	\end{equation}
	with $\pi_1\left(w\right)$ given by \eqref{eq:pi1-triv-case}. Since the map $w\mapsto \pi_1\left(w\right)+\pi_2\left(\mathcal{L}_G\left(w\right)\right)$ is compact, by means of the Fredholm's Alternative Theorem we obtain that $\Sp{\mathcal{L}_G}\subset\Vp{\mathcal{L}_G}\cup\{-d_1\}$.
	
	Take now $w\in\cont$ eigenvector of $\mathcal{L}_G$ with associated eigenvalue $\lambda$, carrying this into equation \eqref{eq:L_G2} we obtain
	 \begin{equation*}
	\frac{\lambda+d_1}{-d_2}\,w=\pi_1\left(w\right)+\lambda\,\pi_2\left(w\right).
	\end{equation*}
	then $w\in \hzero$ and Equation \eqref{eq:eigenvectors-L_G} follows from this last relation and Eqs. \eqref{eq:pi1-triv-case} and \eqref{eq:def:pi2}.
\end{proof}
\end{lemma}
\begin{theorem}\label{theo:stability-sans-inertie-U-fixe} For every $\mathbf{U}\in \mathbb{R}^2$ there exists $\epsilon=\epsilon\left(\mathbf{U}\right)>0$ such that if $\norm{h^+}{\infty}<\epsilon$, then the solution $\left(R_s,p_s\right)$ of problem \eqref{eq:problem-stationary} is asymptotically stable for the evolution problem \eqref{eq:abstract-reynolds}-\eqref{eq:abstract-rp-inertialess}-\eqref{eq:bc-rp-inertialess}.
	\begin{proof}
		Let us assume first that $h^+=0$. By Lemma \ref{lemma:spec-L_G} it is enough to study the eigenvalues of $\mathcal{L}_G$. Hence, take $\lambda\in\mathbb{C}\setminus \{-d_1\}$ such that $\mathcal{L}_G\left(w\right)=\lambda w$ for some $w\neq 0$. Then \eqref{eq:eigenvectors-L_G} reads
		\begin{align}
		h_0^2\,d_3\left(d_1+\lambda\right)\Delta w=&d_4\,\mathbf{U}\cdot \nabla w+\lambda \,d_5\,w&&\hspace*{-2.8cm}\textnormal{in } \Omega,\label{eq:eigenvectors-L_G-simplified}\\
		w=&0 &&\hspace*{-2.8cm}\textnormal{on } \partial \Omega.\nonumber
		\end{align}
		
		We notice that $\lambda\neq 0$. In fact, if $\lambda=0$ then multiplying the Eq. \eqref{eq:eigenvectors-L_G-simplified} by $w$ and integrating by parts we obtain $w=0$, which is not possible. Decomposing $\lambda=\lambda_1+i\,\lambda_2$ and $w=w_1+i\,w_2$, writing the differential equation for the real and imaginary parts we obtain the equations
		\begin{align*}
	-h_0^2\,d_3\left(d_1+\lambda_1\right)\Delta w_1+h_0^2\,d_3\lambda_2\Delta w_2+d_4 \mathbf{U}\cdot\nabla w_1+d_5\left(\lambda_1 w_1-\lambda_2w_2\right)&=0,\\
		-h_0^2\,d_3\left(d_1+\lambda_1\right)\Delta w_2-h_0^2\,d_3\lambda_2\Delta w_1+d_4\mathbf{U}\cdot\nabla w_2+d_5\left(\lambda_1 w_2+\lambda_2w_1\right)&=0.
		\end{align*} 
		 Multiplying the first equation by $w_1$, the second equation by $w_2$ and integrating by parts we may obtain
		\begin{align*}
		h_0^2\,d_3\left(d_1+\lambda_1\right)\int_\Omega \left|\nabla w_1\right|^2\,d\Omega- h_0^2\,d_3\lambda_2\int_\Omega \nabla w_2 \nabla w_1\,d\Omega+d_5\lambda_1\int_\Omega  w_1^2\,d\Omega-d_5\lambda_2\int_\Omega w_1\,w_2\,d\Omega&=0,\\
		h_0^2\,d_3\left(d_1+\lambda_1\right)\int_\Omega \left|\nabla w_2\right|^2\,d\Omega+ h_0^2\,d_3\lambda_2\int_\Omega \nabla w_2 \nabla w_1\,d\Omega+d_5\lambda_1\int_\Omega  w_2^2\,d\Omega+d_5\lambda_2\int_\Omega w_1\,w_2\,d\Omega&=0.
		\end{align*} 
		Adding up both equations we have
		$$h_0^2\,d_3\left(d_1+\lambda_1\right)\int_\Omega \left(\left|\nabla w_1\right|^2+\left|\nabla w_2\right|^2\right)d\Omega+d_5\lambda_1\int_\Omega \left( \left| w_1\right|^2+\left| w_2\right|^2\right)d\Omega=0.$$
		Observing that $\lambda_1\geq 0$ implies $w=0$, which
		 is not possible, we conclude that $\mbox{Re}\left(\lambda\right)=\lambda_1<0$.
		 
		 We have shown the stability for $h^+=0$. Now from Theorem \ref{theo:existence_U_fix} we have that the mapping $h^+\mapsto R_s\left(h^+\right)$ is continuous in a neighborhood $V_1\ni 0$ in $L^\infty\left(\Omega\right)$. Thus, if $h^+\rightarrow 0$ in $L^\infty\left(\Omega\right)$ then $$\norm{DG\left(R_s\left(h^+\right),0\right)-DG\left(\bar{R},0\right)}{}\rightarrow 0$$ in the space of linear continuous operators from $\cont$ into itself. Then the result follows from Proposition \ref{prop:continuity-spec}.
	\end{proof}
\end{theorem}

\begin{theorem} \label{theo:stability-sans-inertie-h-fix}
	 Fix $h\in B_{m_0,M_0}$, $0<m_0<M_0$. Then there exists $\epsilon>0$ such that if $\norm{\mathbf{U}}{}<\epsilon$ then the solution $\left(R_s,p_s\right)$ of problem \eqref{eq:problem-stationary} is asymptotically stable for the evolution problem \eqref{eq:abstract-reynolds}-\eqref{eq:abstract-rp-inertialess}-\eqref{eq:bc-rp-inertialess}.
	\begin{proof}
		Let us assume first that $\mathbf{U}=0$. By Lemma \ref{lemma:spec-L_G} it is enough to study the eigenvalues of $\mathcal{L}_G$. Hence, take $\lambda\in\mathbb{C}\setminus \{-d_1\}$ such that $\mathcal{L}_G\left(w\right)=\lambda w$ for some $w\neq 0$. If $\lambda=0$ then from Eq. \eqref{eq:eigenvectors-L_G} we obtain $w=0$, which is a contradiction. Thus, we have $\lambda\neq 0$ and this time Eq. \eqref{eq:eigenvectors-L_G} in its variational version reads
		\begin{align*}
		\begin{split}
		-d_3\frac{\left(\lambda+d_1\right)}{\lambda}\int_\Omega h^3\nabla w\nabla \phi\,d\Omega&= \, d_5\int_\Omega hw\phi\,d\Omega\qquad \forall \phi\in \hzero.
		\end{split}
		\end{align*}
		Along the same arguments used in Theorem \ref{theo:stabilite-avec-inertie-h-fixe} this implies $\lambda\in\mathbb{R}^-$.
		The result follows analogously to the end of Theorem  \ref{theo:stability-sans-inertie-U-fixe} proof, this time using the continuity of the mapping $\mathbf{U}\mapsto R_s\left(\mathbf{U}\right)$ asserted in Theorem \ref{theo:existence_h_fix}.
	\end{proof}
\end{theorem}

{\color{\coloremph}\begin{remark}
	Theorem \ref{theo:stability-sans-inertie-U-fixe} highlights the difference between the model without or with inertial terms. If $h^+=0$, stability is proved by Theorem \ref{theo:stability-sans-inertie-U-fixe} for any velocity $\mathbf{U}$, while instability is gained for $\mathbf{U}$ sufficiently large from Theorem \ref{theo:instabilite-avec-inertie-U-fixe}.
\end{remark}}

\section{Numerical examples}
\label{sec:numerical-analysis}
In this section we show some numerical examples for the evolution problem \eqref{eq:abstract-reynolds}-\eqref{eq:abstract-rp-inertialess}-\eqref{eq:bc-rp-inertialess}. The numerical method employed consists in a Finite Volume Method to discretize Eq. \eqref{eq:abstract-reynolds} and a backward Euler scheme to discretize Eq. \eqref{eq:abstract-rp-inertialess}. For more details on the numerical method the reader is referred to \cite{Jaramillo2019}.
\begin{figure}[h!]\centering
	\includegraphics[width=0.6\linewidth]{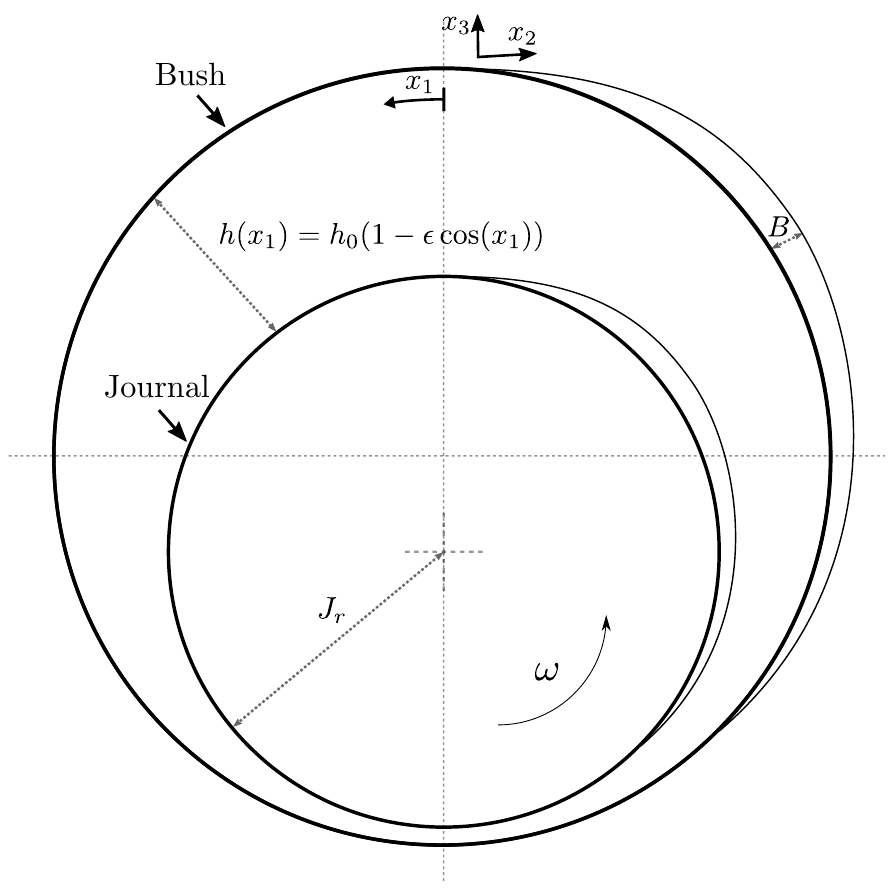}
	\caption{Scheme of the Journal Bearing.}\label{fig:journal-bearing-scheme}
\end{figure}
 The domain $\Omega=[0,2\pi J_r]\times [0,B]$ is divided into cells of size $\Delta x_1 = 2\pi J_r / 512$ and $\Delta x_2 = B/ 64$ in the $x_1$ and $x_2$ axis respectively, where $J_R=B=25.4\times 10^{-3}$ m and. The time step was taken as $\delta t=3\times 10^{-4}$ s. The number of time steps, denoted by $N^*$, is taken big enough in order to observe temporal convergence for each case. Thus, $t^{N^*}=\delta t \,N^*$ corresponds to the final time simulated.
The gap function $h$ was set as $h(x_1,x_2)=h_0\,(1-\epsilon\,\cos(x_1))$ with $h_0>0$ and $\epsilon\in [0,1[$ is the eccentricity. Dirichlet boundary conditions are imposed, reading
$$p\left(x_1,0\right)=p(x_1,B)=0\qquad \forall x_1\in [0,2\pi J_R],$$
and the next periodic conditions
$$p\left(0,x_2\right)=p(2\pi J_R,x_2),~\parder{p}{x_1}\left(0,x_2\right)=\parder{p}{x_1}(2\pi J_R,x_2)\qquad\forall x_2\in [0,B].$$
The initial conditions are $\hat{R}(x_1,x_2,t=0)=R(x_1,x_2,t=0)/R_0=1$ and $\parder{R}{t}=0$ in $\Omega$.

Here the gas fraction is written as (see Appendix)
\begin{equation}
\alpha\left(R\right) = \frac{\alpha_0 \left(R/R_0\right)^3}{1+\alpha_0\left(R/R_0\right)^3}\qquad\forall R>0,\label{eq:def:alpha}
\end{equation}
where $\alpha_0$ is a data corresponding to the gas fraction for $R=R_0$, and $R_0$ is a reference radius.

The geometrical setting corresponds to a journal bearing device, which scheme is shown in Fig. \ref{fig:journal-bearing-scheme}. The physical parameters setting is given in Table \ref{tab:parameters-journal}. For the next results we will use the non-dimensional variables $\hat{R}=R/R_0$, $\hat{x}_1=x_1/J_R$ (longitudinal direction) and $\hat{x}_2=x_2/B$ (transverse direction) {\color{\coloremph} and for some function $f(R)$ we denote $\hat{f}(\hat{R})=f(R_0\,\hat{R})$}.
\begin{table}[h!]
	\begin{center}
		\begin{tabular}{llll}
			\toprule 
			Symbol & Value & Units & Description  \\ 
			\midrule
			$\rho_\ell$ & 854 & kg/m$^3$ & Liquid density \\ 
			$\mu_\ell$ & $7.1\times 10^{-3}$ & Pa$\cdot$s & Liquid viscosity\\
			$\rho_g$ & $1$ &  kg/m$^3$ & Gas density\\
			$\mu_g$ & $1.81\times 10^{-5}$ & Pa$\cdot$s & Gas viscosity\\ 
			$\kappa^s$ & $7.85\times 10^{-5}$ & Pa$\cdot$s$\cdot$m & Surface dilatational viscosity\\ 
			$k$ & $1.4$ &  & Gas polytropic exponent\\
			$\sigma$ & $3.5\times 10^{-2}$ &N/m& Liquid surface tension \\
			$P_0$ & $1$ & atm& Reference pressure\\
			$p_\partial$ & $1$ & atm& Pressure at the boundary\\
			$R_0$ & $3.85\times 10^{-7}$ & m& Bubbles' equilibrium radius at 1 atm\\
			$\alpha_0$ & $0.1$ &  & Reference gas fraction \\ 
			$J_r$ & $25.4\times 10^{-3}$ & m & Journal radius \\
			$B$ & $25.4\times 10^{-3}$ & m & Journal width \\ 
			$h_0$ & $0.001\times J_r$ & m & Journal clearance \\
			$\epsilon$ & $[0,1[$ & & Journal eccentricity\\
			$\omega$ & $2\pi\frac{1000}{60}$  & rad/s & Journal rotational speed\\
			\bottomrule
		\end{tabular} 
		\caption{Parameter values for the Journal Bearing.}\label{tab:parameters-journal}
	\end{center}
\end{table}
\subsubsection*{Time-convergence towards a stationary solution} 
\begin{figure}[h!]\centering
	\includegraphics[width=1.0\linewidth]{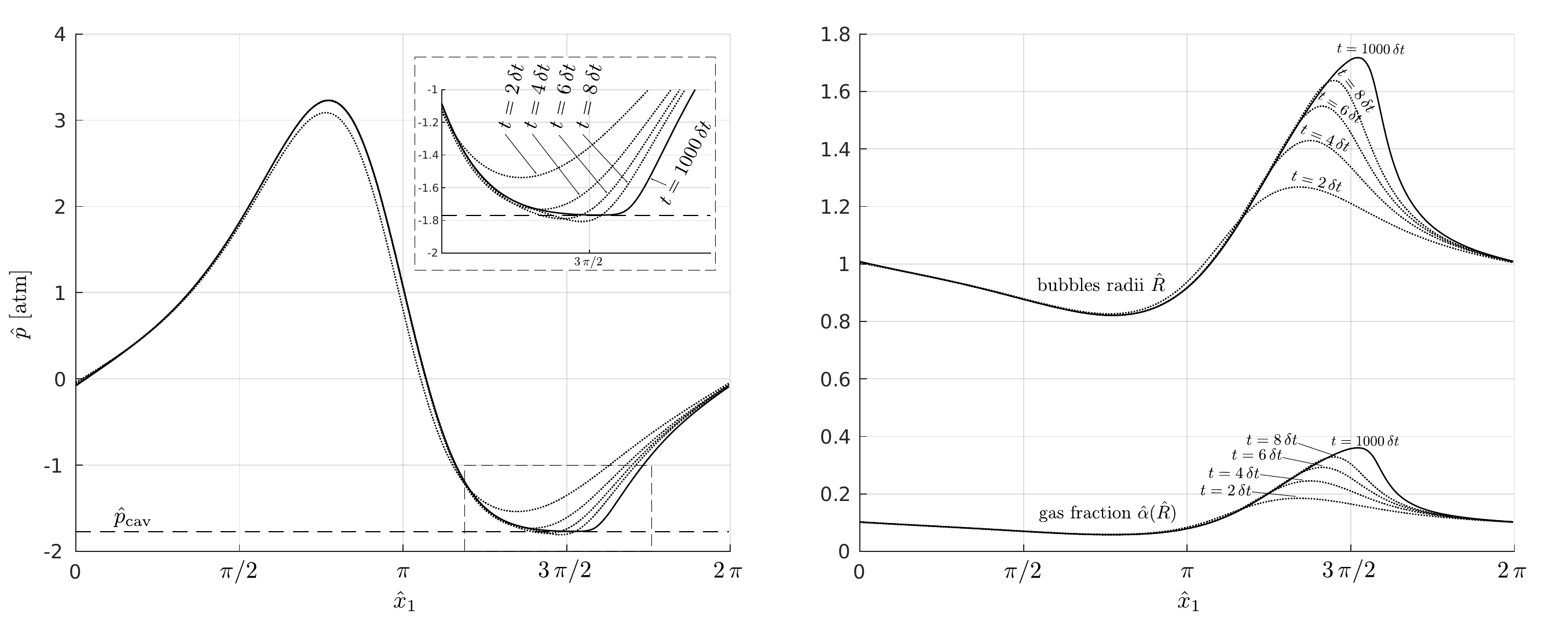}
	\caption{Time convergence of the fields pressure (left), and bubbles dimensionless radii and gas fraction (right) for $\epsilon=0.4$ projected along $\hat{x}_2=0.5$.}\label{fig:time-convergences}
\end{figure}

{ \color{\coloremph}
Here the Journal eccentricity is fixed to $\epsilon=0.4$ and the other physical parameters are set as in Table \ref{tab:parameters-journal}. For the physical cases computed here the function $\hat{f}_1$ has a unique critical point $\hat{R}_\textnormal{crit}$ such that $\hat{f}_1'(R)<0$ for $\hat{R}<\hat{R}_\textnormal{crit}$ and $\hat{f}_1'(R)>0$ for $\hat{R}>\hat{R}_\textnormal{crit}$ (see Fig. \ref{fig:f1}), so we denote $$\hat{p}_{\footnotesize\mbox{cav}}=\min_{r>0}\,\hat{f}_1\left(r\right)=\hat{f}_1(\hat{R}_{\footnotesize\mbox{crit}}).$$
\begin{figure}[h!]\centering
	\includegraphics[width=0.6\linewidth]{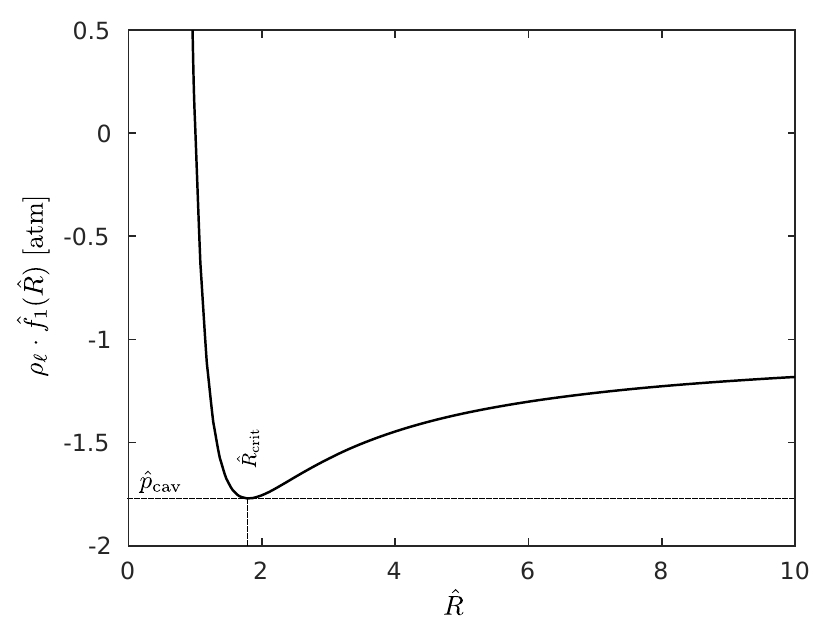}
	\caption{Typical shape of $\hat{f}_1$, having a unique critical point $\hat{R}_\textnormal{crit}\approx 1.8$ for the parameters set in Table \ref{tab:parameters-journal}.}\label{fig:f1}
\end{figure}

To simplify the exposition, these two-dimensional pressure and bubbles' radii fields are shown in Fig. \ref{fig:time-convergences} by fixing $\hat{x}_2=0.5$ for different time-step. For the configuration set in this example a numerical convergence in time is obtained, meaning that for $t\geq 1000\,\delta t$ the profiles are not observed to change. It is worth noticing that for some time steps ($t\approx 7 \delta t$) there is a region of $\Omega$ where $\hat{p}<\hat{p}_\textnormal{cav}$ but the converged profile accomplishes $\hat{p}(\cdot,1000\,\delta t)\geq \hat{p}_\textnormal{cav}$ on $\Omega$. Also, one observes that in the \emph{pressurized} region (where $\hat{p}>0$) the bubbles radii is such that $\hat{\alpha}(\hat{R})$ is low and then $\hat{\rho}(\hat{R}) \approx \rho_\ell$. On the other hand, in the region where $\hat{p}\approx \hat{p}_\textnormal{cav}$ the gas fraction $\hat{\alpha}(\hat{R})$ can reach values as high as $0.4$, lowering the mixture average density and effective viscosity (see (\ref{eq:def-rhoalpha}) and (\ref{eq:def-mualpha})). }

\subsubsection*{Stationary solutions varying the eccentricity}
\begin{figure}[h!]\centering
	\includegraphics[width=1.0\linewidth]{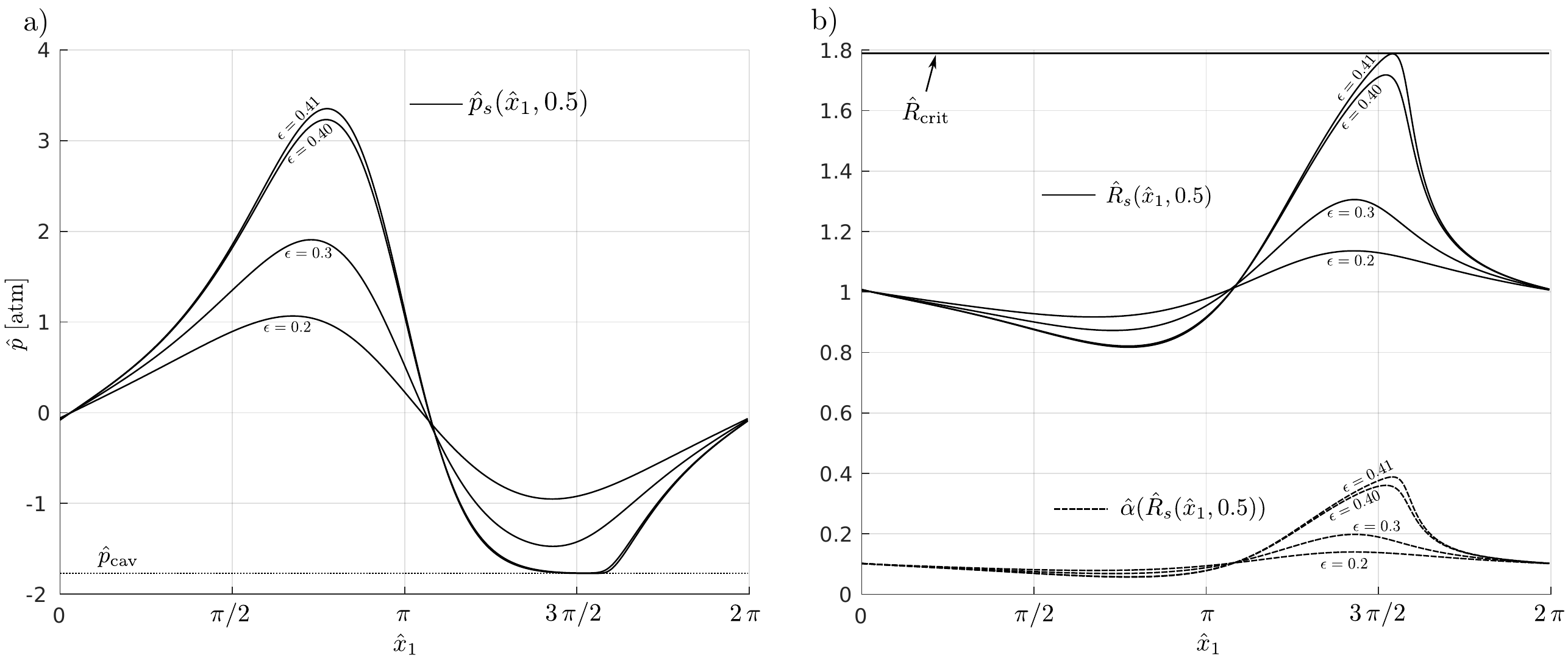}
	\caption{Stationary pressure (left), dimensionless radii and gas-fraction (right) for different eccentricities of the Journal Bearing.}\label{fig:stationary-fields}
\end{figure}

{\color{\coloremph}A series of simulations were performed for increasing values of the eccentricity $\epsilon$. Until a value of $\epsilon$ around $0.41$ time convergence of the transient solution towards the stationary one is numerically observed, while for $\epsilon>0.41$ the time-convergence towards a stationary solution is no longer obtained. Let us remark that the same loss of time convergence is numerically observed when increasing the Journal rotational speed $\omega$.
	
As can be observed from Fig. \ref{fig:stationary-fields} b), the maximum value of $\hat{R}$ on the domain increases as the eccentricity increases, reaching the value $\hat{R}_\textnormal{crit}$ for $\epsilon\approx 0.41$. Thus, the loss of time convergence could be related to the change in the sign of $\hat{f}_1'$ from negative to positive and then violating hypothesis (H1), which was essential to obtain the stability of the stationary solutions in Section \ref{sec:stability-wo-inertia}.}

\section{Future work}
\label{sec:future-work}
Many questions related to the present work could be source of future research, among them:
\begin{itemize}
	\item To include bubbles convection by setting a non-null convective field $\mathbf{u}_b$. This could allow to consider more realistic physical settings, for instance taking $\mathbf{{u}}_b$ equal to the velocity profile corresponding to the Reynolds equation, as done in some numerical works \cite{snyder2016,Jaramillo2019,braun2017}.
	\item Nowadays, it is unclear which are the physical configurations for which the inertial terms on the Rayleigh-Plesset equation are negligible. Numerical works would give a better insight into this. Also, up to our knowledge, there is a lack of robust numerical methods to perform simulations for the full problem.
\end{itemize}

\clearpage
\section*{Appendix}\label{sec:appendix}

Here heuristic arguments to justify Eqs. \eqref{eq:rayleigh-plesset-full} and \eqref{eq:Reynolds} are presented. This presentation is based in the Theory of Multicomponent Fluids \cite{Drew1999}, which has been used by Carrica and coworkers to study the dynamics of a bubbly mixture around a surface ship \cite{Carrica1998,Carrica1999,Carrica2013,Carrica2016}.

\subsection*{The mass and momentum conservation equations}
 The fluid mixture is composed by two phases: an incompressible liquid phase (with density $\rho_\ell$ and viscosity $\mu_\ell$) and a gas phase (with reference density $\rho_g$ and viscosity $\mu_g$). The mixture density and velocity vector are denoted by $\rho$ and $\mathbf{u}$ and the pressure field is denoted by $p$. The characteristic functions of the phases are denoted by $X_k(\mathbf{x},t)$, $k=\ell,g$ for the liquid and gas phase respectively. This is, $X_k(\mathbf{x},t)=1$ if the phase $k$ is present in $\mathbf{x}$ at time $t$, and $X_k(\mathbf{x},t)=0$ otherwise. 

An ensemble of physical realizations is assumed to exist. Each of these realizations corresponds to an evolution of the physical system (fluid mixture - limiting surfaces - boundary conditions) for which the initial conditions are \emph{near enough} to a set of ideal smooth initial conditions. A probability of occurrence for each realization is assumed. Here $\langle \cdot \rangle$ denotes the statistical averaging with respect to this distribution function at the point $\mathbf{x}\in\Omega^V$ and time $t$. The fields 
\begin{equation}
\alpha_g\left(\mathbf{x},t\right) = \av{X_g},\qquad \alpha_\ell(\mathbf{x},t)=\av{ X_\ell},
\end{equation}
are called the \emph{gas fraction}, and the \emph{liquid fraction} respectively. In the statistical model, the phases occupy the whole domain, thus $\alpha_g+\alpha_\ell=1$.  With this, the averaged density, velocity vector and stress tensor for each phase $k=\ell,g$ are defined as
\begin{equation}
\bar{\rho}_k(\mathbf{x},t) = \frac{\av{\rho X_k} }{\alpha_k},\qquad \mathbf{\bar{u}}_k(\mathbf{x},t) = \frac{\av{\rho \mathbf{{u}}\, X_k} }{\alpha_k \bar{\rho}_k},\qquad \mathbf{\bar{T}}_k=\frac{\av{ \mathbf{T}\,X_k}}{\alpha_k}.
\end{equation}

Observe that as the liquid phase is incompressible and so one has that $\bar{\rho}_\ell=\rho_\ell$. Applying the averaging process to the conservation equations of mass and momentum, one obtains \cite{Drew1999}
\begin{equation}
\parder{(\alpha_k\bar{\rho}_k)}{t}+\nabla\cdot \left(\alpha_k\bar{\rho}_k\,\mathbf{\bar{u}}_k\right)=\Gamma_k\qquad\mbox{in }\Omega^V,
\label{eq:av-mass-cons0}
\end{equation}
and
\begin{equation}
\parder{(\alpha_k\bar{\rho}_k\,\mathbf{u}_k)}{t}+\nabla\cdot\left(\alpha_k\bar{\rho}_k\mathbf{\bar{u}}_k\otimes \mathbf{\bar{u}}_k\right)= \nabla\cdot\left(\alpha_k\,\mathbf{\bar{T}}_k\right)+\mathbf{M}_k+v_{ki}\,\Gamma_k\qquad\mbox{in }\Omega^V,\label{eq:av-momentum-cons0}
\end{equation}
where $\mathbf{M}_k$ is the interfacial momentum source and, $v_{ki}$ is the interfaces speed and $\Gamma_k$ is the interfacial mass generation source. In the literature, a series of hypotheses are made to simplify these equations:

\begin{itemize}
	\item Averaged quantities are smooth;
	\item Interfacial mass sources are negligible ($\Gamma_k\simeq 0$);
	\item Interfacial momentum sources are negligible ($\mathbf{M}_k\simeq 0$);
	\item The gas phase average velocity is equal to the liquid phase average velocity, $\mathbf{\bar{u}}_g=\mathbf{\bar{u}}_\ell$;
	\item The gas phase density is equal to the gas reference density $\rho_g$, $\bar{\rho}_g=\rho_g$.
\end{itemize}
With these hypotheses, adding up Eq. \eqref{eq:av-mass-cons0} for both phases, simplifying the notation by setting $\alpha\overset{\scriptsize\mbox{def}}{=}\alpha_g$, and introducing the variable
\begin{equation}
\bar{\rho}\left(\alpha\right)=\alpha\left(\mathbf{x},t\right)\rho_\ell+\left(1-\alpha\left(\mathbf{x},t\right)\right)\rho_g,
\label{eq:def-rhoalpha}
\end{equation}
it is obtained the conservation law:
\begin{equation}
\parder{\bar{\rho}}{t}+\nabla\cdot \left(\bar{\rho}\,\mathbf{\bar{u}}_\ell\right)=0 \qquad\textnormal{in }\Omega^V. \label{eq:av-mass-cons1}
\end{equation}
Similarly, adding up Eq. \eqref{eq:av-momentum-cons0} for both phases we have
\begin{equation}
\parder{(\bar{\rho}\,\mathbf{\bar{u}}_\ell)}{t}+\nabla\cdot\left(\bar{\rho}\,\mathbf{\bar{u}}_\ell\otimes \mathbf{\bar{u}}_\ell\right)=-\nabla \bar{p}+\mu_{\scriptsize{\textnormal{eff}}}\,\nabla^2 \mathbf{\bar{u}}_\ell \qquad\textnormal{in }\Omega^V,
\label{eq:av-momentum-cons1}
\end{equation}
where it has been assumed the existence of an effective fluid viscosity $\mu_{\scriptsize{\textnormal{eff}}}$ (that depends smoothly on $\alpha$, the liquid viscosity and the gas viscosity) such that
\begin{equation}
\nabla\cdot\left(\alpha\,\mathbf{\bar{T}}_g+(1-\alpha)\mathbf{\bar{T}}_\ell\right)\simeq-\nabla \bar{p}+\mueff\,\nabla^2 \bar{\mathbf{u}}_\ell \qquad\textnormal{in }\Omega^V.
\label{eq:tensor-app}
\end{equation}
An example off $\mueff$ is given by \cite{Saadat1996}
\begin{equation}
\mueff(\alpha)=\alpha\left(\mathbf{x},t\right) \mu_g + (1-\alpha\left(\mathbf{x},t\right)) \mu_\ell.\label{eq:def-mualpha}
\end{equation}
To our knowledge, there is a lack of works justifying \eqref{eq:tensor-app} or some similar relation, and further research on the topic is needed.

The thin film hypothesis allows to approximate the Navier-Stokes Eqs. \eqref{eq:av-momentum-cons1} along Eq. \eqref{eq:av-mass-cons1} by the incompressible Reynolds equation \cite{Cameron1971,Bayada1986}:
\begin{equation}
\nabla_x\cdot \left(\frac{\bar{\rho} h^3}{12\mueff}\,\nabla \bar{p}\right)=\nabla_x\cdot \left(\frac{\mathbf{U}}{2}\,\bar{\rho} h\,\right)+\parder{\bar{\rho}h}{t}\qquad\textnormal{in }\Omega.\label{eq:ReynoldsAp}
\end{equation} 

\subsection*{The field Rayleigh-Plesset equation}
In the literature regarding the Rayleigh-Plesset equation to model the evolution of bubbly fluids the following hypotheses are generally made:
\begin{itemize}
\item The liquid phase of the mixture is continuous while the gas phase corresponds to a high number of spherical bubbles dispersed in the liquid \cite{natsumeda1987,Carrica1998,Carrica1999,Carrica2013,snyder2016,Carrica2016,Jaramillo2019};
\item The bubbles remains small enough and the distance between them remains large enough in such a way that the pressure gradients are locally negligible;
\item The flow of the liquid phase is radially symmetrical around each bubble;
\item The bubbles radii distribution is monodisperse, i.e., if $f(r,\mathbf{x},t)$ is the distribution such that for each $\mathbf{x}$ and $t$ the number of bubbles of size between $r$ and $r+dr$ is $f(r,\mathbf{x},t)\,dr$, and $n_b(\mathbf{x},t)$ is the number of bubbles per unit volume then there exists a field $R$ depending only on $\mathbf{x}$ and $t$ such that
$$f(r,\mathbf{x},t)=n_b(\mathbf{x},t)\,\delta(r-R(\mathbf{x},t));$$
\item The evolution of the field $R\left(\mathbf{x},t\right)$ is related to the average liquid pressure $\bar{p}_\ell$ by means of the Rayleigh-Plesset equation
\begin{equation}
\rho_\ell\left[\frac{3}{2}\left(\parderD{R}{t}\right)^2+R\parderDtwo{R}{t}\right]=P_0\left(\frac{R_0}{R}\right)^{3k}-\left(\bar{p}_\ell+p_\partial\right)-\frac{2\sigma}{R}-4\left(\frac{\mu_\ell+\kappa^s/R}{R}\right)\parderD{R}{t},\label{eq:rayleigh-plesset-fullAp}
\end{equation}
where $\rho_\ell$ and $\rho_g$ ($\mu_\ell$ and $\mu_g$) are the densities (viscosities) of the liquid and the gas respectively, $P_0$ is the inner pressure of the bubble when its radius is equal to $R_0$, $k$ is the polytropic exponent, $\sigma$ is the surface tension, $\kappa^s$ is the surface dilatational viscosity \cite{snyder2016} and $p_\partial$ is the pressure at the boundary.
Regarding the material derivative
\begin{equation}
\parderD{}{t} = \parder{}{t}+\left(\mathbf{{u}}_b\cdot \nabla\right),\label{eq:Dt_appendix}
\end{equation}
the convective field $\mathbf{{u}}_b=(u_b^1,u_b^2,u_b^3)$ is three-dimensional and when coupled to the Reynolds equation, it is generally assumed that $u_b^3=0$ (e.g., \cite{Jaramillophd}). Thus, the physics of the thin-film are three-dimensional but the mathematical modeling by means of Eqs. \eqref{eq:rayleigh-plesset-fullAp} and \eqref{eq:ReynoldsAp} is two-dimensional.
\end{itemize}

 {\color{\coloremph}In general, the energy equation must be considered in this kind of modeling. However, in this work the polytropic exponent is set to $k=1$ or $1.4$ (air specific heat), the former value corresponds to an isothermal process while the latter one corresponds to an adiabatic process, both cases where the energy equation is not needed. Nevertheless, it is worth noticing that there exist related works where the energy equation is also considered (e.g., \cite{braun2017, Meng2016}).}

It can be proved by that the three variables $n_b\left(\mathbf{x},t\right)$, $\alpha\left(\mathbf{x},t\right)$ and $R\left(\mathbf{x},t\right)$ accomplish the geometric relation (e.g., \cite{Drew1999} Section 10.1.2)
\begin{equation}
\alpha\left(\mathbf{x},t\right)=n_b\left(\mathbf{x},t\right)\frac{4\pi R\left(\mathbf{x},t\right)^3}{3}.\label{eq:alphaR}
\end{equation}
To couple \eqref{eq:rayleigh-plesset-fullAp} with Eq. \eqref{eq:ReynoldsAp} there remains to give and additional model for $n_b\left(\mathbf{x},t\right)$. Two approaches may be found in the literature:
\begin{enumerate}
	\item To assume that $n_b\left(\mathbf{x},t\right)$ is a known constant.
	\item To assume that the number of bubbles per unit of liquid, denoted $n_b^\ell$, is constant (e.g., \cite{zwart2004}) and so
	$$n_b\left(\mathbf{x},t\right)=n_b^\ell\, \alpha_\ell\left(\mathbf{x},t\right) = n_b^\ell (1-\alpha\left(\mathbf{x},t\right))$$
	that combined with Eq. \eqref{eq:alphaR} implies
	\begin{equation}
	\alpha\left(\mathbf{x},t\right)=\frac{n_b^\ell\frac{4\pi R\left(\mathbf{x},t\right)^3}{3}}{1+n_b^\ell\frac{4\pi R\left(\mathbf{x},t\right)^3}{3}}.\label{eq:alphanb}
	\end{equation}
	
	Notice that this expression is bounded by the unit and it grows monotonically with $R$.
	In the literature it is typically introduced the parameter $\alpha_0=n_b^\ell\frac{4\pi}{3}R_0^3$ which corresponds to a reference gas fraction. Doing so, this last equation may be written
	\begin{equation*}
	\alpha\left(R\right) = \frac{\alpha_0 \left(R/R_0\right)^3}{1+\alpha_0\left(R/R_0\right)^3},
	\end{equation*}
	that corresponds to the formula used in Section \ref{sec:numerical-analysis} for the numerical examples.
\end{enumerate}
The boundedness of $\alpha$ obtained from the second approach is one of the hypotheses made in Section \ref{sec:math-framework}. Thus, the theoretical results proved in this work remain valid for other definitions of $\alpha\left(R\right)$ accomplishing that property.

If a \emph{polydisperse} distribution of bubbles is assumed, one may adapt the multigroup approach used by Carrica and coworkers for the modeling on the interaction of ocean air bubbles with a surface ship \cite{Carrica1999,Carrica2013,Carrica2016}, where a Population Balance Equation is written and a discrete group of possible bubbles radii is assumed. The use of this methodology for the context of the Reynolds-Rayleigh-Plesset cavitation model is an ongoing research topic.

\section*{Acknowledgments}

The authors thank the financial support of this work provided by CAPES (grant PROEX-8434433/D); {\em O primeiro autor foi bolsista
da CAPES PDSE Processo n$^\circ$ 88881.133497/2016-01}.

\bibliographystyle{unsrt}

\begin{thebibliography}{10}
	
	\bibitem{dowson1979}
	D.~Dowson and C.~Taylor.
	\newblock {Cavitation in bearings}.
	\newblock {\em {Annu. Rev. Fluid Mech.}}, 11:35--66, 1979.
	
	\bibitem{braun2010}
	M.~Braun and W.~Hannon.
	\newblock {Cavitation formation and modelling for fluid film bearings: {A}
		review}.
	\newblock {\em {J. Engineering Tribol.}}, 224:839--863, 2010.
	
	\bibitem{Cimatti1976}
	G.~Cimatti.
	\newblock On a problem of the theory of lubrication governed by a variational
	inequality.
	\newblock {\em Applied Mathematics and Optimization}, 3(2):227--242, 1976.
	
	\bibitem{Cimatti1980}
	G.~Cimatti.
	\newblock A free boundary problem in the theory of lubrication.
	\newblock {\em International Journal of Engineering Science}, 18(5):703--711,
	1980.
	
	\bibitem{kinderlehrer1980}
	D.~Kinderlehrer and G.~Stampacchia.
	\newblock {\em {An Introduction to Variational Inequalities and Their
			Applications}}, chapter VII, pages 223--227.
	\newblock {SIAM}, 1980.
	
	\bibitem{bayada1983}
	G.~Bayada and M.~Chambat.
	\newblock Analysis of a free boundary problem in partial lubrication.
	\newblock {\em Quart. Appl. Math.}, 40(4):369--375, 1983.
	
	\bibitem{Bermudez1989}
	A.~Berm\'udez and J~Durany.
	\newblock Numerical solution of cavitation problems in lubrication.
	\newblock {\em Computer Methods in Applied Mechanics and Engineering},
	75(1):457--466, 1989.
	
	\bibitem{buscaglia2015a}
	G.~C. Buscaglia, Mohamed {El Alaoui Talibi}, and M.~Jai.
	\newblock {Mass-conserving cavitation model for dynamical lubrication problems.
		Part I: Mathematical analysis}.
	\newblock {\em Mathematics and Computers in Simulation}, 118:130--145, 2015.
	
	\bibitem{Bayada2006}
	G.~Bayada, S.~Martin, and C.~V\'azquez.
	\newblock {About a generalized Buckley-Leverett equation and lubrication
		multifluid flow}.
	\newblock {\em Euro. Jnl. of Applied Mathematics}, 17(5):491--524, 2006.
	
	\bibitem{schnerr2001}
	G.~H. Schnerr and J.~Sauer.
	\newblock {Physical and numerical modeling of unsteady cavitation dynamics}.
	\newblock In {\em {Fourth International Conference on Multiphase Flow}}, 2001.
	
	\bibitem{singhal2001}
	A.~K. Singhal, H.~Y. Li, M.~M. Athavale, and Y.~Jiang.
	\newblock {Mathematical Basis and Validation of the Full Cavitation Model}.
	\newblock {\em {J. Fluids Eng}}, 124(3):617--624, 2001.
	
	\bibitem{zwart2004}
	P.~J. Zwart, A.~G. Gerber, and T.~Belarmi.
	\newblock {A two-phase flow model for predicting cavitation dynamics}.
	\newblock In {\em {Fifth International Conference on Multiphase Flow}}, 2004.
	
	\bibitem{kawase1985}
	T.~Kawase and T.~Someya.
	\newblock Investigation into the oil film pressure distribution in dynamically
	loaded journal bearings.
	\newblock {\em Transactions of the Japan Society of Mechanical Engineers Series
		C}, 51(470):2562--2570, 1985.
	
	\bibitem{natsumeda1987}
	S.~Natsumeda and T.~Someya.
	\newblock Paper iii(ii) negative pressures in statically and dynamically loaded
	journal bearings.
	\newblock In {\em Fluid Film Lubrication - Osborne Reynolds Centenary},
	volume~11 of {\em Tribology Series}, pages 65 -- 72. Elsevier, 1987.
	
	\bibitem{kubota1992}
	A.~Kubota, H.~Kato, and H.~Yamaguchi.
	\newblock A new modelling of cavitating flows: a numerical study of unsteady
	cavitation on a hydrofoil section.
	\newblock {\em Journal of Fluid Mechanics}, 240:59--96, 1992.
	
	\bibitem{gehannin2009}
	J.~Gehannin, M.~Arghir, and O.~Bonneau.
	\newblock {Evaluation of Rayleigh-Plesset Equation Based Cavitation Models for
		Squeeze Film Dampers}.
	\newblock {\em Journal of Tribology}, 131(2):024501, 2009.
	
	\bibitem{geike2009a}
	T.~Geike and V.~Popov.
	\newblock {A bubble dynamics based approach to the simulation of cavitation in
		lubricated contacts}.
	\newblock {\em Journal of Tribology}, 131(1):011704, 2009.
	
	\bibitem{geike2009b}
	T.~Geike and V.~Popov.
	\newblock {Cavitation within the framework of reduced description of mixed
		lubrication}.
	\newblock {\em Tribology International}, 42(1):93--98, 2009.
	
	\bibitem{Jaramillo2019}
	A.~Jaramillo and G.~Buscaglia.
	\newblock {A stable numerical strategy for Reynolds-Rayleigh-Plesset coupling}.
	\newblock {\em {Tribol. Int.}}, 130:191--205, 2019.
	
	\bibitem{Jaramillophd}
	A.~Jaramillo.
	\newblock {\em New models and Numerical Methods in Hydrodynamic Lubrication}.
	\newblock PhD thesis, Instituto de Ci\^encias Matem\'aticas e de
	Computa\c{c}\~ao, USP, 2019.
	
	\bibitem{snyder2016}
	T.~A. Snyder, M.~J. Braun, and K.~C. Pierson.
	\newblock {Two-way coupled Reynolds and Rayleigh-Plesset equations for a fully
		transient, multiphysics cavitation model with pseudo-cavitation}.
	\newblock {\em Tribology International}, 93:429--445, jan 2016.
	
	\bibitem{brennen1995}
	C.~E. Brennen.
	\newblock {\em {Cavitation and Bubble Dynamics}}.
	\newblock Oxford University Press, 1995.
	
	\bibitem{Gehannin2016}
	J.~Gehannin, M.~Arghir, and O.~Bonneau.
	\newblock A volume of fluid method for air ingestion in squeeze film dampers.
	\newblock {\em Tribology Transactions}, 59(2):208--218, 2016.
	
	\bibitem{Drew1999}
	D.~Drew and S.~Passman.
	\newblock {\em {Theory of Multicomponent Fluids}}.
	\newblock Springer-Verlag New York, 1999.
	
	\bibitem{liuzzi2012}
	D.~Liuzzi.
	\newblock {\em {Two-Phase Cavitation Modelling}}.
	\newblock PhD thesis, University of Rome, 2012.
	
	\bibitem{schmidt2014}
	M.~Schmidt, P.~Reinke, M.~Nobis, and M.~Riedel.
	\newblock {Three-dimensional simulation of cavitating flow in real journal
		bearing geometry}.
	\newblock In {\em 4th Micro and Nano Flows Conference (MNF2014)}, pages 7--10,
	2014.
	
	\bibitem{liu2014}
	Y.~Liu, L.~Wang, and Z.~Zhu.
	\newblock {Numerical study on flow characteristics of rotor pumps including
		cavitation}.
	\newblock {\em Proceedings of the Institution of Mechanical Engineers, Part C:
		Journal of Mechanical Engineering Science}, 229(14):2626--2638, 2015.
	
	\bibitem{walters2015}
	M.~J. Walters.
	\newblock {\em {An Investigation into the Effects of Viscoelasticity on
			Cavitation Bubble Dynamics with Applications to Biomedicine}}.
	\newblock PhD thesis, Cardiff University, 2015.
	
	\bibitem{zhao2016}
	Y.~Zhao, G.~Wang, and B.~Huang.
	\newblock {A cavitation model for computations of unsteady cavitating flows}.
	\newblock {\em Acta Mechanica Sinica}, 32(2):273--283, 2016.
	
	\bibitem{dhande2016}
	D.~Y. Dhande and D.~W. Pande.
	\newblock Multiphase flow analysis of hydrodynamic journal bearing using cfd
	coupled fluid structure interaction considering cavitation.
	\newblock {\em Journal of King Saud University - Engineering Sciences}, 2016.
	
	\bibitem{hakl2012}
	R.~Hakl, P.~J. Torres, and M.~Zamora.
	\newblock Periodic solutions to singular second order differential equations:
	the repulsive case.
	\newblock {\em Topol. Methods Nonlinear Anal.}, 39(2):199--220, 2012.
	
	\bibitem{ohnawa2016}
	M.~Ohnawa and Y.~Suzuki.
	\newblock {Mathematical and Numerical Analysis of the Rayleigh-Plesset and the
		Keller Equations}.
	\newblock In {\em Mathematical Fluid Dynamics, Present and Future}, pages
	159--180. Springer Japan, 2016.
	
	\bibitem{papanicolaou1978}
	A.~Bensoussan, J.~L. Lions, and G.~Papanicolaou.
	\newblock {\em Asymptotic Analysis for Periodic Structures}, chapter~1,
	page~38.
	\newblock North-Holland, 1978.
	
	\bibitem{lang1999}
	Lang S.
	\newblock {\em Fundamentals of Differential Geometry}, chapter~I, pages 19--20.
	\newblock Springer-Verlag New York, 1999.
	
	\bibitem{dunford1957}
	N.~Dunford and J.~T. Schwartz.
	\newblock {\em Linear Operators, Part I: General Theory}, chapter VII, page
	585.
	\newblock Interscience Publishers, 1957.
	
	\bibitem{Benzoni2010}
	S.~Benzoni-Gavage.
	\newblock {\em {Calcul diff\'erentiel et \'equations diff\'erentielles: cours
			et exercices corrigés}}.
	\newblock {SIAM}, 2010.
	
	\bibitem{Gantmacher1959}
	Gantmacher~F. R.
	\newblock {\em Applications of the Theory of Matrices}, chapter~V, page 230.
	\newblock Interscience Publishers, 1959.
	
	\bibitem{braun2017}
	M.~Braun, K.~Pierson, and T.~Snyder.
	\newblock {Two-way coupled Reynolds, Rayleigh-Plesset-Scriven and energy
		equations for fully transient cavitation and heat transfer modeling}.
	\newblock {\em IOP Conference Series: Materials Science and Engineering},
	174:012030, 2017.
	
	\bibitem{Carrica1998}
	P.~M. Carrica, F.~J. Bonetto, D.~A. Drew, and R.~T. Lahey.
	\newblock {The interaction of background ocean air bubbles with a surface
		ship}.
	\newblock {\em International Journal for Numerical Methods in Fluids},
	28(4):571--600, 1998.
	
	\bibitem{Carrica1999}
	P.~Carrica, D.~Drew, F.~Bonetto, and R.~Lahey.
	\newblock A polydisperse model for bubbly two-phase flow around a surface ship.
	\newblock {\em Int. J. Multiphase Flow}, 25:257--305, 1999.
	
	\bibitem{Carrica2013}
	A.~Castro and P.~Carrica.
	\newblock Bubble size distribution prediction for large-scale ship flows: Model
	evaluation and numerical issues.
	\newblock {\em Int. J. Multiphase Flow}, 57:131--150, 2013.
	
	\bibitem{Carrica2016}
	A.~Castro, J.~Li, and P.~Carrica.
	\newblock A mechanistic model of bubble entrainment in turbulent free surface
	flows.
	\newblock {\em Int. J. Multiphase Flow}, 86:35--55, 2016.
	
	\bibitem{Saadat1996}
	N.~Saadat and W.~Flint.
	\newblock {Expressions for the Viscosity of Liquid/Vapour Mixtures: Predicted
		and Measured Pressure Distributions in a Hydrostatic Bearing}.
	\newblock {\em {Proc. Inst. Mech. Eng. J. J. Eng. Tribol.}}, 210(1):75--79,
	1996.
	
	\bibitem{Cameron1971}
	A.~Cameron.
	\newblock {\em {Basic Lubrication Theory}}.
	\newblock Longman Publishing Group Ltd., 1971.
	
	\bibitem{Bayada1986}
	G.~Bayada and M.~Chambat.
	\newblock {The transition between the {Stokes} equations and the {Reynolds}
		equation: A mathematical proof}.
	\newblock {\em {Applied Mathematics \& Optimization}}, 14(1):73--93, 1986.
	
	\bibitem{Meng2016}
	{Effect of Groove Textures on the Performances of Gaseous Bubble in the
		Lubricant of Journal Bearing}.
	\newblock {\em Journal of Tribology}, 139(3):031701, 2016.
	
\end{thebibliography}

\end{document}